\documentclass[aps,prd,onecolumn,groupedaddress,nofootinbib,amssymb]{revtex4}
\usepackage{bm}

\usepackage{amsmath}
\usepackage{amssymb}
\usepackage{amsfonts}
\usepackage{wrapfig}
\usepackage{cancel} 
\usepackage[dvipdfmx]{graphicx}
\usepackage{color}
\newcommand{\be}{\begin{equation}}
\newcommand{\ee}{\end{equation}}
\newcommand{\bea}{\begin{eqnarray}}
\newcommand{\eea}{\end{eqnarray}}
\newcommand{\beaa}{\begin{eqnarray*}}
\newcommand{\eeaa}{\end{eqnarray*}}

\newcommand{\ms}{\mathcal{S}}
\allowdisplaybreaks[4]

\begin{document}

\title{Covariant constraints of massive gravity in metric formulation}

\author{Satoshi Akagi$^{1}$\footnote{
	E-mail address: akagi.satoshi@b.mbox.nagoya-u.ac.jp} and Taisaku Mori$^{1}$\footnote{
	E-mail address: mori.taisaku@k.mbox.nagoya-u.ac.jp}}

\affiliation{
$^1$ Department of Physics, Nagoya University, Nagoya
464-8602, Japan}

\begin{abstract}
We propose a simple method for deriving the constraints of the de Rham-Gabadadze-Tolley model in the metric and the Lagrangian formulation, as possible as keeping the Lorentz covariance. In our formulation, it is not necessary to use the Hamilton analysis, the vielbein formulation, nor the St\"uckelberg trick for showing the Boulwer-Deser ghost-freeness. 
It realizes the Lorentz covariant expressions of the constraints in a certain parameter region.

\end{abstract}

%\pacs{}

\maketitle

\section{INTRODUCTION}
In a last decade, the understanding of massive spin-two field has been quite developed.
The unique linear theory describing the massive spin-two particle in the flat spacetime has been well-known since a long time ago. It is called the Fierz-Pauli (FP) model \cite{Fierz:1939ix}.
The equations of motion (EoM) of the FP model can be expressed as follows,
\begin{align}
(\Box-m^2) h_{\mu\nu} =0, \ \ \partial^\mu h_{\mu \nu}-\partial_\nu h =0, \ \ h=0. \label{inin1}
\end{align}
Although the second equations can be rewritten as $\partial^\mu h_{\mu\nu}=0$ by using the third equation $h=0$, we adopt this expression in (\ref{inin1}) for a {later} convenience. 
In the Fourier space, the first equation determines a dispersion relation, and the other equations can be regarded as the constraints which reduce the components of the polarizations.
Hence, in the $D$ dimensional spacetime, the symmetric tensor $h_{\mu\nu}$ obeying the FP model has the $(D+1)(D-2)/2$ components.
In four dimensions, {the FP model has five degrees of freedom (DoF)}, and it coincides with the DoF of the spin-two representation of the little group $SO(3)$. Indeed, in four dimensions, the wave equations (\ref{inin1}) can be derived from the representation theory of the Poincare group $ISO(1,3)$. Hence, (\ref{inin1}) can be regarded as a definition of the wave equation for the symmetric second rank tensor describing the massive spin-two particle.

In 1970's, from a negative standpoint, it began to ask whether or not the gravity can be described by the massive spin two particle.
It had been pointed out that the original FP model cannot explain the observational results in the solar system \cite{vanDam:1970vg}.
This is because the massless limit of some observables in the FP model {does} not coincide with {those} of the massless model. This discontinuity is called the vDVZ discontinuity.
Although it seems that the possibility of the non-vanishing graviton mass had been excluded from the above fact, 
Vainshtein proposed that a class of nonlinearlization of the FP model can avoid the vDVZ discontinuity \cite{Vainshtein:1972sx}.
%some extensions of the FP model, such as non-linearizations or curved space extensions, are not straightforward.
%In particular, a class of the nonlinearization called the massive gravity has been quite developed in a last decade.
In order to clarify the {latter} explanation, we would like to define the ``massive gravity" generally as the sum of the Einstein-Hilbert action with dynamical metric $g_{\mu\nu}$ and non-derivative potential terms which are general functions of the dynamical metric $g_{\mu\nu}$ and the flat metric $\eta_{\mu\nu}$, i.e.,
\begin{align}
S= M^{D-2}_g \int d^Dx \sqrt{-g}\left[R(g)  - V(g,\eta)\right]. \label{inin2}
\end{align}
Here, we assume a general nonderivative potential term $\sqrt{-g}V(g,\eta)$ which is expandable {with respect to} $h_{\mu\nu}=g_{\mu\nu}-\eta_{\mu\nu}$. 
In the expansion, we assume also that the quadratic order terms with respect to $h_{\mu\nu}$ are proportional to the Fierz-Pauli mass term $h^2 - h^{\mu\nu}h_{\mu\nu}$ and the first order terms vanish.
 %Furthermore, we assume the $V(g,h)$ which is Lorenz covariant.  
Then, the action (\ref{inin2}) can be regarded as a nonlinear extension of the FP model in flat spacetime. 
We should note that there are no general covariance in the action (\ref{inin2}).
But we assume that the $V(g,\eta)$ which {has} the Lorentz covariance.

In \cite{Vainshtein:1972sx}, Vainshtein considered the model (\ref{inin2}) where the potential term $\sqrt{-g}V(g,\eta)$ is taken to be {the} FP mass term, and he {has shown} that there are no discontinuities in the massless limit. 
%consider the massive gravity (\ref{inin2}) in 1970's in order to explain the observational results in the solar system \cite{vanDam:1970vg,Vainshtein:1972sx}.
However, the Hamiltonian corresponding to the general action (\ref{inin2}), without any special tunings, is unbounded and contains a ghost-like mode because of the violation of a constraint corresponding to $h=0$ in (\ref{inin1}). This ghost mode is called the the Boulware-Deser (BD) ghost \cite{Boulware:1974sr}.
Although the BD ghost problem had been remaining during a long time, in 2010, the BD ghost problem had been solved by discovering the 
de Rham-Gabadadze-Tolley (dRGT) model \cite{deRham:2010ik,deRham:2010kj}.
In \cite{deRham:2010ik}, de Rham and Gabadadze  tried to tune the parameters in the massive gravity model (\ref{inin2}) by requiring the ghost-freeness in the high-energy limit called the decoupling limit. 
And they concluded that {the Lorentz covariant potential term $V(g,\eta)$ consistent with this requirement is parametrized by only three free parameters (mass parameter, and other two dimensionless parameters) in $D=4$.} Although the analysis in \cite{deRham:2010ik} was perturbative in powers of $h_{\mu\nu}$, in \cite{deRham:2010kj}, the full-nonlinear form of the action was obtained. 
After that, the BD ghost-freeness of the dRGT model without taking any limits was proved by using the Hamilton analysis \cite{Hassan:2011hr,Hassan:2011ea} with the Arnowitt-Deser-Misner (ADM) variables \cite{ADM}.
Furthermore, as an extension of the dRGT model, it was pointed out that the model {where} the flat metric $\eta_{\mu\nu}$ {is replaced by} an arbitrary fixed metric $f_{\mu\nu}$ called the fiducial metric is also ghost-free \cite{Hassan:2011tf}.

Let us summarize known results on the constraint analysis of the dRGT model.
As we have mentioned, the first works in \cite{deRham:2010ik,deRham:2010kj} argue the BD ghost-freeness in the decoupling limit by using the so called St\"uckelberg  trick.
And the proof of the full order model in \cite{Hassan:2011hr,Hassan:2011ea,Hassan:2011tf} is based on the Hamilton analysis.
Nowadays, in the metric formulation, there are a lot of works \cite{kluson1,kluson2,kluson3,kluson4,golovnev,mirbabayi,HassanS,comelli,kugo1}  based on the St\"uckelberg trick and/or the Hamilton analysis with the ADM variables.
On the other hand, in the {vielbein} description of the dRGT model \cite{viel1}, there are some works based on the Lagrangian formulation \cite{viel3,DeserS}.
These works derive the constraints as possible as keeping the ``Lorentz covariance" (for $f_{\mu\nu}=\eta_{\mu\nu}$).
Thus. the analysis becomes quite tractable and realize the Lorentz covariant expressions of the constraints in a certain parameter region. 
%In this formulation, the analysis is simplified since we can keep the original variables.
However, it has still been unknown how to derive the covariant expressions of the constraints (covariant constraints) in the metric formulation except for the linearized case \cite{bernard1,bernard3,bernard2}.

Our purposes is to derive the constraints with the metric and the Lagrangian formulation, as possible as keeping the Lorentz covariance.
It can be predicted that this approach {could} realize the covariant expressions of the constraints.
%to derive the covariant constraints corresponding to the constraints in (\ref{inin1}).
{The difficulty in the case of the metric formulation comes from how to treat the square root matrix $S^\mu_{~~\nu} \equiv \sqrt{g^{-1}f}^\mu_{~~\nu}$ which construct{s} the potential term{s} in the dRGT model.}
Fortunately, there have been some works for the model linearlized around a general background solution \cite{bernard1,bernard3,bernard2}. From the results of the linearlized model, we can guess which {of} the linear combinations of the EoM realize the covariant constraints in the full-nonlinear level.
Although the method in \cite{bernard1,bernard3,bernard2} cannot straightforwardly be extended to the nonlinear case,
we can avoid this difficulty by making a good use of the algebraic relation,
\begin{align}
\nabla_\lambda \ms^{\mu}_{~~\nu} \ms^{\nu}_{~~\rho} + \ms^{\mu}_{~~\nu} \nabla_\lambda \ms^{\nu}_{~~\rho}
= g^{\mu\nu} \nabla_\lambda f_{\nu \rho},
\end{align}
{for} the square root matrix $\ms^\mu_{~~\nu} = \sqrt{g^{-1}f}^\mu_{~~\nu}$.
As the results, we derive the explicit form{s} of the covariant constraints in a certain parameter region.
Furthermore, for the general parameter region, we prove the existence of the constraints corresponding to the constraints in (\ref{inin1}).

This paper is organized as follows:
In Sec.\ref{dr}, we summarize the fundamental properties of the dRGT model.
In Sec.\ref{lin}, we consider the model linearized around a general background solution and obtain some suggestions for deriving the covariant constraints of the nonlinear model.  
In Sec.\ref{cov}, we derive the explicit form of the covariant constraints for a certain parameter region.
In Sec.\ref{cons} and Sec.\ref{pro}, we prove the existence of the constraint{s} for the general case in any dimensions. 
In Sec\ref{cons}, in order to make the understanding on the constraint structure clear,
we demonstrate that the existence of the constraint{s} becomes trivial if we admit an identity.
In Sec.\ref{pro}, we prove the identity used in Sec.\ref{cons}.
{The last section \ref{summary} is devoted to the summary}.

{Let us summarize the notations in this paper.
The bracket $( \mu_1 \mu_2 \cdots \mu_n )$ denotes the totally symmetrization for the inner indexes $\mu_1 \mu_2 \cdots \mu_n $ with the weight $1/n!$, e.g., $A^{(\mu_1 \mu_2)} \equiv \frac{1}{2!} \left( A^{\mu_1 \mu_2} + A^{\mu_2 \mu_1}\right)$.
Similarly, the bracket $[\mu_1 \mu_2 \cdots \mu_n]$ denotes the totally anti-symmetrization for the inner indexes $\mu_1 \mu_2 \cdots \mu_n $ with the weight $1/n!$, e.g., $A^{[\mu_1 \mu_2]} \equiv \frac{1}{2!} \left(A^{\mu_1 \mu_2} - A^{\mu_2 \mu_1}\right)$.
Furthermore, the covariant derivative is defined by $\nabla_\mu A_{\nu} \equiv \partial_{\mu} A_{\nu} - \Gamma_{\mu\nu}^{\lambda} A_{\lambda}$.
The Riemann curvature is given by $[\nabla_\mu , \nabla_\nu] A_{\sigma} \equiv R_{\mu\nu\sigma \rho} A^\rho$.
Using the Riemann curvature, the Ricci curvature can be expressed as $R_{\mu\nu} \equiv R_{\mu_1 \mu \nu_1 \nu}g^{\mu_1 \nu_1}$.}

\section{dRGT model}
\label{dr}
Let us start with summarizing the fundamental properties of the dRGT model.
In order to make the argument as possible as general, we consider the dRGT model with an arbitrary fiducial metric $f_{\mu\nu}$,
\begin{align}
&S_{\text{dRGT}}= M_g^{D-2} \int d^D x\sqrt{-g}  \left[ R(g) 
-2\sum_{n=0}^{D-1} \beta_n e_{(n)} (\mathcal{S}) \right], \notag \\
&\mathcal{S}^\mu_{ \ \ \nu} \equiv {\sqrt{g^{-1}f}}^\mu_{ \ \ \nu }.
\label{int2}
\end{align}
Here,  the symmetric polynomial $e_{(n)}(\ms)$ is defined as follows,
\begin{align}
e_{(0)} (\ms) &\equiv1, \notag \\
e_{(1)}(\ms) &\equiv \ms^\mu_{~\mu}, \notag \\
e_{(n)}(\ms) &\equiv  
 \ms^{\mu_1}_{ \ \ [\mu_1} \ms^{\mu_2}_{ \ \ \mu_2} \cdots \ms^{\mu_n }_{ \ \ \mu_n]}  \notag \\
& =\frac{1}{n!} g^{\mu_1 \nu_1 \mu_2 \nu_2 \cdots \mu_n \nu_n} \ms_{\mu_1 \nu_1} \ms_{\mu_2 \nu_2} 
\cdots \ms_{\mu_n \nu_n} \ \ (n=2,3,\cdots, D-1) \label{ly2}. 
\end{align}
In the last line, we lower the index of $\ms^\mu_{~\nu}$ by using the metric as $g_{\mu\rho}\ms^\rho_{~~\nu}\equiv \ms_{\mu\nu}$ and define the higher rank tensor $g^{\mu_1 \nu_1 \mu_2 \nu_2 \cdots \mu_n \nu_n}$ as a totally antisymmetrization of the product $g^{\mu_1 \nu_1} g^{\mu_2 \nu_2} \cdots g^{\mu_n \nu_n}$ with respect to $\nu_1, \ \nu_2, \cdots \nu_n$, i.e.,
\begin{align}
g^{\mu_1 \nu_1 \mu_2 \nu_2 \cdots \mu_n \nu_n}
&\equiv  n! \delta^{\nu_1}_{[\rho_1} \delta^{\nu_2}_{\rho_2} \cdots \delta^{\nu_n}_{\rho_n]}
g^{\mu_1 \rho_1} g^{\mu_2 \rho_2} \cdots g^{\mu_n \rho_n}.
\end{align}
Here, we eliminate the weight factor from the definition of $g^{\mu_1 \nu_1 \cdots \mu_n \nu_n}$, e.g., $g^{\mu_1 \nu_1 \mu_2 \nu_2} \equiv g^{\mu_1 \nu_1} g^{\mu_2 \nu_2} - g^{\mu_1 \nu_2} g^{\mu_2 \nu_1}$. The fundamental properties of this tensor is summarized in Appendix \ref{ap1}. Furthermore, $\beta_n$ denote the free parameters. Although the mass parameter $m^2$ is commonly introduced by replacing $\beta_n \rightarrow m^2 \beta_n$ in (\ref{int2}),
we omit this mass parameter in order to regard $\beta_n$ as the free parameters.

And the square root matrix $\ms^\mu_{~~\nu}\equiv \sqrt{g^{-1}f}^\mu_{~~\nu}$ is defined so that it satisfies the relation,
\begin{align}
\ms^\mu_{~~\rho} \ms^\rho_{~~\nu}= g^{\mu\rho} f_{\rho\nu}.\label{ly1}
\end{align}
{We assume Det$(f_{\mu\nu})\neq0$ and {therefore}, Det$(\ms^\mu_{~~\nu})\neq0$.}
Furthermore, we should note the following important relation that $\ms_{\mu\nu} \equiv g_{\mu \rho} \ms^\rho_{~\nu}$ is symmetric with respect to $\mu\nu$,\footnote{
This symmetric property can be shown from the following relation, 
\begin{align}
\sqrt{g^{-1}f}^\mu_{~\nu}=g^{\mu\rho} \sqrt{f g^{-1}}_\rho^{~\sigma} g_{\sigma \nu}, \label{ftn1}
\end{align}
which can be confirmed by expanding the square root of the quantity $\sqrt{g^{-1}f}^{\mu}_{~\nu}=\sqrt{1+(g^{-1}f-1)}^\mu_{~\nu}$ (for example see \cite{Hassan4}).}
\begin{align}
\ms_{\mu\nu} = \ms_{\nu\mu}. \label{ftn2}
\end{align}

Our purpose is to derive the constraints of the dRGT model described by the action (\ref{int2}).
In the system where the highest order time derivative terms in the EoM is second order, the constraints can be defined as the linear combinations of the EoM{ which do not include} any second order time derivatives of the dynamical variables, i.e., $\partial_0 \partial_0 g_{\mu_1 \nu_1}$.\footnote{The validity of this definition of the constraints is explained by the Lagrangian analysis (see \cite{Buchbinder1,new curved,BTM} for the case of the FP model).}
Then, {we now} derive the EoM of the action (\ref{int2}).

A nontrivial point is how to calculate the variation of the symmetric polynomial, $\delta e_{(n)}(\ms)$.
From the definition (\ref{ly2}), $\delta e_{(n)}(\ms)$ is given by,
\begin{align}
\delta e_{(n)}(\ms)  =
 {Y_{(n-1)}}_{\mu}^{~~\nu}(\ms)  \delta \ms^{\mu}_{~~\nu},\label{oy1}
\end{align}
where the matrix $Y^{\mu\nu}_{(n)}(\mathcal{S})$ is defied as follows,
\begin{align}
Y^{\mu\nu}_{(n)}(\mathcal{S}) \equiv \frac{1}{n!} g^{\mu\nu\mu_1 \nu_1 \mu_2 \nu_2 \cdots \mu_n \nu_n} \mathcal{S}_{\mu_1 \nu_1}
\mathcal{S}_{\mu_2 \nu_2} \cdots \mathcal{S}_{\mu_n \nu_n}.
\end{align}
We should note that the symmetric property $Y^{\mu\nu}_{(n)}(\ms)=Y^{(\mu\nu)}_{(n)}(\ms)$ {follows} the symmetry {(\ref{ftn2})} and the symmetric properties (\ref{Appp1}) given in the Appendix \ref{ap1}.
In order to calculate the right hand side of (\ref{oy1}), we {use} the following algebraic relation,
\begin{align}
\delta \ms^{\mu}_{~~\nu} \ms^{\nu}_{~~\rho} + \ms^{\mu}_{~~\nu} \delta \ms^{\nu}_{~~\rho}
=\delta g^{\mu\nu} f_{\nu \rho}, \label{li4}
\end{align}
which is obtained by variating the relation (\ref{ly1}).
By {multiplying} the matrix $\ms^{-1\rho}_{~~~~~\sigma} {Y_{(n)}}^\sigma_{~~\mu}(\ms)$ on the both sides of the above relation (\ref{li4}) and using the commutativity $\left[ Y_{(n)}(\ms), \ms  \right]=0$, we obtain the relation, $\delta \ms^{\mu}_{~~\nu}  {Y_{(n)}}^{\nu}_{~~\mu} (\ms) 
=\frac12 \delta g^{\mu\rho} \ms_{\rho \nu}  {Y_{(n)}}_{~~\mu}^{\nu} (\ms)$.
By substituting this relation, $\delta \ms^{\mu}_{~~\nu}  {Y_{(n)}}^{\nu}_{~~\mu} (\ms) 
=\frac12 \delta g^{\mu\rho} \ms_{\rho \nu}  {Y_{(n)}}_{~~\mu}^{\nu} (\ms)$, into (\ref{oy1}), we obtain,
\begin{align}
\delta e_{(n)}(\ms) = \frac12 \delta g^{\mu\rho} \ms_{\rho \nu}  {Y_{(n-1)}}_{~~\mu}^{\nu} (\ms).
\label{oy2}
\end{align}
Using Eq.(\ref{oy2}) and the recursion relation, $Y^{\mu\nu}_{(n)}(\ms) = g^{\mu\nu}e_{(n)}(\ms) - \ms^\mu_{~\rho} Y^{\rho \nu}_{(n-1)} (\ms)$, given in Eq.(\ref{ly3}) of Appendix \ref{ap1}, 
we find,
\begin{align}
\delta \left[ \sqrt{-g} e_{(n)}(\ms) \right] 
=-\frac{\sqrt{-g}}{2} Y_{(n)\mu\nu}(\ms) \delta g^{\mu\nu}.
\end{align}
Hence, the EoM corresponding to the action (\ref{int2}) can be determined as follows,
\begin{align}
E^{\mu\nu} \equiv  G^{\mu\nu} + \sum_{n=0}^{D-1} \beta_n Y_{(n)}^{\mu\nu} (\ms) =0. \label{ly5}
\end{align}
Here, we define the Einstein tensor by $G^{\mu\nu}\equiv R^{\mu\nu} -Rg^{\mu\nu}/2$.

As we have mentioned, in the system where the highest order time derivative terms in the EoM is second order, the constraints can be defined as the linear combinations of the EoM independent of any second order time derivatives of the dynamical metric, $\partial_0 \partial_0 g_{\mu_1 \nu_1}$.
One of the constraints is given by taking the divergence of the EoM (\ref{ly5}) and using the Bianchi identity $\nabla_\mu G^{\mu\nu}=0$,
\begin{align}
\nabla_\mu E^{\mu\nu} = \sum_{n=1}^{D-1} \beta_n \nabla_\mu Y_{(n)}^{\mu\nu} (\ms) =0.\label{by1}
\end{align}
Obviously, this equation is independent of $\partial_0 \partial_0 g_{\mu_1 \nu_1}$. 
Hence, (\ref{by1}) can be regarded as a constraint corresponding to $\partial^\mu h_{\mu\nu} -\partial_\nu h= 0$ in (\ref{inin1}).
{We should note that the special tuning of the potential term in the dRGT model (\ref{int2}) is not essential for the existence of the constraint (\ref{by1}).
In other words,
for a model whose potential term is different from that of the dRGT model, such as $V(g,\eta)$ in (\ref{inin2}),
the constraint corresponding to (\ref{by1}) exists.}

The special tuning of the potential terms in the dRGT model (\ref{int2}) {is essential for} the existence of an additional constraint corresponding to $h=0$ in (\ref{inin1}).
The covariant expression of this constraint has not been obtained and it is the purpose in this paper.

\section{Suggestions from linearized model}
\label{lin}
The purpose of this section is to obtain a few suggestions for deriving the covariant constraints in the nonlinear model by investigating the covariant constraints of the model linearized around 
a general background solution. Just for simplicity, in this section, we consider only the case of $\beta_0\neq0, \ \beta_1\neq0, \ \beta_n =0 \ (n=2,3,\cdots ,D-1)$.
Fortunately, for the linearized model, the linear combination expressing the additional constraint has been obtained in \cite{bernard1,bernard3,bernard2}.
In the case of $\beta_0\neq0, \ \beta_1\neq0, \ \beta_n =0 \ (n=2,3,\cdots ,D-1)$, by denoting the linearized EoM as $\delta E^\mu_{~~\nu}=0$, the following linear combination of $\delta E^{\mu}_{~~\nu}$ does not contain any second order derivative of the perturbation $\delta g_{\mu\nu}$,
\begin{align}
 \delta \phi_1 \equiv \nabla_\nu \left( \ms^{-1\nu\rho} \nabla_\mu \delta E^{\mu}_{~~\rho}\right)
+ \frac{\beta_1}{D-2}\delta E^{\rho}_{~~\rho}. \label{li3}
\end{align}
However, the method in \cite{bernard1,bernard3,bernard2}  cannot straightforwardly be extended to the nonlinear case. Then, in this section, we show {that Eq.(\ref{li3}) does not include} $\partial_\mu \partial_\nu \delta g_{\mu_1 \nu_1}$ or equivalently, $\delta \phi_1=0$ is a constraint, by a different way {which can be extended to the nonlinear case.}

In the case of  $\beta_0\neq0, \ \beta_1\neq0, \ \beta_n =0 \ (n=2,3,\cdots ,D-1)$, the EoM of the nonlinear model given in (\ref{ly5}) is expressed as follows,
\begin{align}
E^{\mu}_{~\rho} \equiv G^{\mu}_{~\rho} + \beta_0 \delta^{\mu}_{\rho} + \beta_1
\left( \delta^{\mu}_{\rho} \mathcal{S}  - \ms^{\mu}_{~~\rho} \right) =0.\label{be1}
\end{align}
Here, we define $\ms\equiv \ms^\rho_{~~\rho}$. In this section, we regard this equation as a background equation.
By denoting the metric perturbation as $\delta g_{\mu\nu}= h_{\mu\nu}$, the linearization of the EoM (\ref{ly5}) around any background solution is given by,
\begin{align}
&\delta E^{\mu}_{~\rho} \equiv \delta G^{\mu}_{~\rho}
+\frac{\beta_1}{2} g_{\rho \nu}\left( g^{\mu (\mu_1} \mathcal{S}^{\nu_1) \nu }-  g^{\nu (\mu_1} \mathcal{S}^{\nu_1) \mu} 
+2 g_{~~~~~\mu_2\nu_2}^{(\mu\nu)} \mathcal{S}^{\mu_2\nu_2,\mu_1 \nu_1}  \right)h_{\mu_1 \nu_1}
 =0, \notag \\
&\mathcal{S}^{\mu~~,\alpha \beta}_{~~\nu} \equiv \frac {\partial \mathcal{S}^{\mu}_{~~\nu}}{\partial g_{\alpha \beta}} ,\notag \\
&\delta G^\rho_{~\nu} = \frac12 g^{\rho \mu}\left[ 2\nabla^\sigma \nabla_{(\mu} h_{\nu) \sigma} - \Box h_{\mu\nu} - \nabla_\mu \nabla_\nu h - \nabla^\alpha \nabla^\beta h_{\alpha \beta} g_{\mu \nu} + \Box h g_{\mu \nu}+ \left( -2\delta_{\mu}^{(\mu_1}  R^{\nu_1)}_{\nu} 
+ g_{\mu\nu} R^{\mu_1 \nu_1} \right) h_{\mu_1 \nu_1} \right]
.\label{li2}
\end{align}
Here, we raise the index of the $\mathcal{S}^{\mu~~,\alpha \beta}_{~~\rho}$ by using $g^{\nu\rho}$ as $\ms^{\mu\nu, \alpha \beta} \equiv g^{\nu \rho} \ms^{\mu~~,\alpha \beta}_{~~\rho} $.
We should note that the superscripts $\mu\nu$ in the $\ms^{\mu\nu, \alpha \beta}$ is {\it not} symmetric with respect to the permutation $\mu\leftrightarrow \nu$ although the matrix $\ms^{\mu}_{~~\rho}g^{\rho\nu} \equiv \ms^{\nu\mu}$ satisfies $\ms^{\mu\nu}=\ms^{(\mu\nu)}$.
The antisymmetric part $\ms^{[\mu\nu], \mu_1 \nu_1}$ is determined as follows,
\begin{align}
\ms^{[\mu\nu], \mu_1 \nu_1} = \frac12 \left( \ms^{\mu(\mu_1} g^{\nu_1) \nu} - \ms^{\nu(\mu_1} g^{\nu_1) \mu} \right), \label{li1}
\end{align}
from the linearization of the relation $\ms^{\mu\nu}=\ms^{\nu\mu}$.
We have used the relation (\ref{li1}) to obtain Eq.(\ref{li2}).

{On the other hand, the symmetric part $\ms^{(\mu\nu),\mu_1 \nu_1}$ is determined from the following relation,
\begin{align}
 \mathcal{S}^{\mu~~,\alpha \beta}_{~~\rho}\mathcal{S}^{\rho}_{~~\nu} 
+\mathcal{S}^{\mu}_{~~\sigma} \mathcal{S}^{\sigma~~,\alpha \beta}_{~~\nu}
= -g^{\mu (\alpha} \left[\mathcal{S}^2\right]^{\beta)}_{~~\nu}, \label{li6}
\end{align}
which is just the relation (\ref{li4}) expressed in terms of $\mathcal{S}^{\mu~~,\alpha \beta}_{~~\nu}$ defined in (\ref{li2}). Here, we define the matrix multiplication $[AB]^{\mu}_{~\nu} \equiv A^\mu_{~\rho}B^{\rho}_{~\nu}$ for any matrices $A^\mu_{~\nu}, \ B^\mu_{~\nu}$.}
In \cite{bernard1,bernard3}, the explicit form of the symmetric part $\ms^{(\mu\nu),\mu_1 \nu_1}$ has been obtained as a function of $\ms^\mu_{~\nu}, \ g_{\mu\nu}$ by solving the relation (\ref{li6}), and the constraint (\ref{li3}) is obtained by using the explicit form of the $\ms^{\mu\nu,\mu_1\nu_1}$.
However, the explicit form of the $\ms^{\mu\nu,\mu_1\nu_1}$ is not so simple and it is not obvious how to extend the linear analysis to the nonlinear case.
Although more tractable method is demonstrated in \cite{bernard2} by redefining the massive spin two field $h_{\mu\nu}$, this method cannot also be extended to the nonlinear case straightforwardly.   
{Then, in this paper, we consider {to show} the existence of the constraint (\ref{li3}), by making a good use of the relation (\ref{li6}), without solving the relation (\ref{li6}) explicitly.}

Let us investigate the constraint structure of the linearized equation (\ref{li2}).
The divergence of Eq.(\ref{li2}), $\nabla_\mu \delta E^{\mu}_{~~\rho}=0$, denotes the vector constraints.
We should note that the divergence of the linearized Einstein tensor $\nabla_\mu \delta G^{\mu}_{~\rho}$ is not equal to zero for general backgrounds because of the non-commutativity of the covariant derivative $\nabla_\mu$ and the variation operator $\delta$. Indeed, for general backgrounds, the linearization of the Bianchi identity $\nabla_\mu G^{\mu}_{~\rho}=0$ becomes,
\begin{align}
0&=\delta (\nabla_\mu G^{\mu}_{~\rho}) \notag \\
&= \nabla_\mu \delta G^{\mu}_{~\rho}  + \frac12g_{\nu\rho} \left( G^{\mu\nu}g^{\mu_1 \nu_1} -g^{\mu\nu}G^{\mu _1 \nu_1}  \right) \nabla_\mu h_{\mu_1 \nu_1}. \label{be4}
\end{align}
We should note that the second line of the above equation (\ref{be4}) becomes equal to zero only in the case of the Einstein manifold where the background metric $g_{\mu\nu}$ satisfies the Einstein equation with a cosmological constraint $\Lambda$ because the Einstein tensor is proportional to metric, i.e{.} $G_{\mu\nu}= \Lambda g_{\mu\nu}$.
In other cases $G_{\mu\nu} \neq \Lambda g_{\mu\nu}$, $\nabla_\mu \delta G^{\mu}_{~\rho}$ is not equal to zero, $\nabla_\mu \delta G^{\mu}_{~~\rho} \neq0$, generally.
Hence, by using the linearized Bianchi identity (\ref{be4}), we obtain,
\begin{align}
g^{\rho\nu }\nabla_\mu \delta E^{\mu}_{~\rho} = &  \frac12 \left(g^{\mu\nu}G^{\mu_1 \nu_1} 
-g^{\mu_1 \nu_1} G^{\mu\nu}\right)\nabla_\mu h_{\mu_1 \nu_1}\notag \\
&+ \frac{\beta_1}{2} \left( g^{\mu(\mu_1} \ms^{\nu_1) \nu} - g^{\nu(\mu_1}\ms^{\nu_1)\mu}  + 2g^{(\mu\nu)}_{~~~~~\mu_2 \nu_2} \ms^{\mu_2 \nu_2, \mu_1 \nu_1}\right) \nabla_\mu h_{\mu_1 \nu_1}\notag \\
&+ (\text{terms without any derivatives of }h) \notag \\
=& \frac{\beta_1}{2} \left( g^{\mu\nu} \mathcal{S}^{\mu_1 \nu_1} -g^{\mu_1 \nu_1}\mathcal{S}^{\mu\nu}  
  + g^{\mu (\nu_1} \mathcal{S}^{\mu_1) \nu }
-\mathcal{S}^{\mu (\nu_1}g^{\mu_1) \nu} +2g^{(\mu\nu)}_{~~~~~\mu_2 \nu_2} \ms^{\mu_2 \nu_2, \mu_1 \nu_1}\right) \nabla_\mu h_{\mu_1 \nu_1} \notag \\
&+ (\text{terms without any derivatives of }h). \label{li5}
\end{align}
In the second line, we use the background equation (\ref{be1}) in order to eliminate the Einstein tensor $G^{\mu\nu}$.
Because this quantity (\ref{li5}) does not contain any second order derivative terms $\partial_\mu \partial_\nu h_{\mu_1 \nu_1}$, under the EoM $\delta E^\mu_{~~\nu}=0$, the relations $\nabla_\mu \delta E^\mu_{~~\nu}=0$ can be regarded as constraints. Those are just the linearization of the vector constraints (\ref{by1}).

Next, let us show that the linear combination (\ref{li3}) does not depend on any second order derivatives of $h_{\mu\nu}$ by using the relation (\ref{li6}).
By multiplying the inverse matrix $\ms^{-1}$ to the vector constraints (\ref{li5}), the quantity $\ms^{-1\nu\rho} \nabla_\mu \delta E^{\mu}_{~~\rho}$ in the linear combination (\ref{li5}) can be expressed as follows,
\begin{align}
\mathcal{S}^{-1\nu}_{~~~~~\rho} g^{\sigma\rho }\nabla_\mu \delta E^{\mu}_{~\sigma}
=& \frac{\beta_1}{2} \left(  -g^{\mu\nu(\mu_1 \nu_1)} 
+ \mathcal{S}^{-1\mu\nu} \mathcal{S}^{\mu_1 \nu_1}
-\mathcal{S}^{\mu (\nu_1}\mathcal{S}^{-1\mu_1) \nu} \right. \notag \\
& \left.+ 2 \mathcal{S}^{-1\mu\nu} 
\mathcal{S}^{\rho~~,\mu_1 \nu_1}_{~~\rho} - 2\mathcal{S}^{-1\nu}_{~~~~~\rho}\mathcal{S}^{(\mu\rho),\mu_1 \nu_1} \right) \nabla_\mu h_{\mu_1 \nu_1} \notag \\
&+ (\text{terms without any derivatives of }h).  \label{li8}
\end{align}
In order to confirm that the linear combination (\ref{li3}) does not depend on $\partial_\mu \partial_\nu h_{\mu_1 \nu_1}$, we have to express the coefficient tensor in front of the second order derivative term $\nabla_\mu \nabla_\nu h_{\mu_1 \nu_1}$ in $\nabla_\nu(\mathcal{S}^{-1\nu}_{~~~~~\rho} g^{\sigma\rho }\nabla_\mu \delta E^{\mu}_{~\sigma})$ of Eq.(\ref{li3}) as in terms of $\ms^{\mu\nu}, \ g_{\mu\nu}$.
Then, we have to deform the tensor $\ms^{\mu\nu,\mu_1 \nu_1}$ in (\ref{li8}) by using the relation (\ref{li6}).
Let us show that it is not necessary to derive the explicit solution of (\ref{li6}) for $\ms^{\mu\nu,\mu_1 \nu_1}$ nor redefine the field $h_{\mu\nu}$ as done in \cite{bernard1,bernard3,bernard2}.
There is a simple way showing that the linear combination (\ref{li3}) does not depend on $\partial_\mu \partial_\nu h_{\mu_1 \nu_1}$.
From the relation (\ref{li6}), we can easily show the following relations,
\begin{align}
& \mathcal{S}^{\rho~~,\mu_1 \nu_1}_{~~\rho} = -\frac12 \mathcal{S}^{\mu_1 \nu_1} ,\label{be11} \\
&{\mathcal{S}^{-1}}^{\mu}_{~~\sigma} \mathcal{S}^{\sigma\nu,\alpha \beta}
+\mathcal{S}^{\mu~~,\alpha \beta}_{~~\rho} \mathcal{S}^{-1\rho\nu}
= -\mathcal{S}^{-1\mu(\alpha} \mathcal{S}^{\beta) \nu}.\label{be10}
\end{align}
We find that Eq.(\ref{be11}) determines the trace part $\mathcal{S}^{\rho~~,\mu_1 \nu_1}_{~~\rho}$ in Eq.(\ref{li8}) as in terms of $\ms^\mu_{~~\nu}$. Furthermore, Eq.(\ref{be10}) determines the symmetrization of $\mathcal{S}^{-1\nu}_{~~~~~\rho}\mathcal{S}^{(\mu\rho),\mu_1 \nu_1} $ with respect to $\mu\nu$.
\if0
\begin{align}
2\mathcal{S}^{-1\nu}_{~~~~~\rho}\mathcal{S}^{(\mu\rho),\mu_1 \nu_1} \supset \mathcal{S}^{-1(\nu}_{~~~~~\rho}\mathcal{S}^{\mu)\rho,\mu_1 \nu_1}
+\mathcal{S}^{-1(\nu|}_{~~~~~\rho}\mathcal{S}^{\rho |\mu),\mu_1 \nu_1}
= -\frac12 \left( \ms^{-1\mu (\mu_1} \ms^{\nu_1) \nu} + \ms^{-1\nu (\mu_1} \ms^{\nu_1) \mu}  \right).
\end{align}
\fi
Although the antisymmetrization of $\mathcal{S}^{-1\nu}_{~~~~~\rho}\mathcal{S}^{(\mu\rho),\mu_1 \nu_1} $ with respect to $\mu \nu$ are not easily determined from the relation (\ref{be10}),
this fact does not affect to our analysis as we will see below.

Then, by using the relations (\ref{be11}) {and} (\ref{be10}), Eq.(\ref{li8}) can be rewritten as follows,
\begin{align}
\mathcal{S}^{-1\nu}_{~~~~~\rho} g^{\sigma\rho }\nabla_\mu \delta E^{\mu}_{~\sigma}
=& -\frac{\beta_1}{2} \left(  g^{\mu\nu(\mu_1 \nu_1)} 
+\mathcal{S}^{\nu_1[\mu }\mathcal{S}^{-1\nu]\mu_1 } 
+  \mathcal{S}^{-1[\nu}_{~~~~~\rho}\mathcal{S}^{\mu] \rho, \mu_1 \nu_1}
+\mathcal{S}^{-1[\nu|}_{~~~~~\rho}\mathcal{S}^{\rho |\mu], \mu_1 \nu_1}
\right) \nabla_\mu h_{\mu_1 \nu_1} \notag \\
&+ (\text{terms without any derivatives of }h). \label{bet2}
\end{align}
Here, the bracket $[ \mu | \cdots | \nu ]$ denotes the antisymmetrization with respect $\mu\nu$, e.g., $A^{[\mu| \lambda | \nu]} \equiv \frac{1}{2}\left( A^{\mu \lambda \nu} - A^{\nu \lambda \mu}\right)$. 
We find the important fact that all the coefficient {matrices} in Eq.(\ref{bet2}) except for $g^{\mu\nu(\mu_1 \nu_1)}$ are antisymmetric with respect to $\mu\nu$.
These terms do not affect to the constraint structure because the covariant derivative in these terms turn into {the} Riemann curvatures by taking the divergence of Eq.(\ref{bet2}), i.e.,
\begin{align}
\nabla_\nu \left(\mathcal{S}^{-1\nu}_{~~~~~\rho} g^{\sigma\rho }\nabla_\mu \delta E^{\mu}_{~\sigma} \right)
=& -\frac{\beta_1}{2}  g^{\mu\nu(\mu_1 \nu_1)} \nabla_\nu \nabla_\mu h_{\mu_1 \nu_1} 
\notag \\
& -\frac{\beta_1}{4} \left( \mathcal{S}^{\nu_1[\mu }\mathcal{S}^{-1\nu]\mu_1 } 
+\mathcal{S}^{-1[\nu}_{~~~~~\rho}\mathcal{S}^{\mu] \rho, \mu_1 \nu_1}
+\mathcal{S}^{-1[\nu|}_{~~~~~\rho}\mathcal{S}^{\rho |\mu], \mu_1 \nu_1}
\right) [\nabla_\nu, \nabla_\mu] h_{\mu_1 \nu_1} \notag \\
&+ (\text{terms without any second derivatives of }h) \notag \\
=&-\frac{\beta_1}{2}  g^{\mu\nu(\mu_1 \nu_1)} \nabla_\nu \nabla_\mu h_{\mu_1 \nu_1} 
+ (\text{terms without any second derivatives of }h).
\end{align}
Then, the only quantity with the second order derivative of $h_{\mu\nu}$ is $g^{\mu\nu(\mu_1 \nu_1)} \nabla_\nu \nabla_\mu h_{\mu_1 \nu_1} $.
However, we can cancel this term by the trace of the EoM (\ref{li2}),
\begin{align}
\delta E^{\rho}_{~\rho} = \frac{D-2}{2} g^{\mu_1 \nu_1 \mu_2 \nu_2} \nabla_{\mu_1}
\nabla_{\nu_1} h_{\mu_2 \nu_2} + (\text{terms without any second derivatives of }h).
\end{align}
Therefore, the following linear combination does not contain any second derivatives of the massive spin-two field,
\begin{align}
\delta \phi_1 \equiv \nabla_\nu \left(\mathcal{S}^{-1\nu}_{~~~~~\rho} g^{\sigma\rho }\nabla_\mu \delta E^{\mu}_{~\sigma} \right)
+\frac{\beta_1}{D-2} \delta E^\rho_{~\rho}.\label{bet3}
\end{align}
Therefore, under the EoM $\delta E^\mu_{~~\nu}=0$, the eqution $\delta \phi_1=0$ can be regarded as a constraint.

\section{Covariant constraints}\label{cov}
From this section, we investigate the additional constraint of the full-nonlinear model described by the EoM (\ref{ly5}). 
In this section, by focusing on the case of  $\beta_0 \neq 0,  \ \beta_1\neq0, \ \beta_2\neq0 \ , \beta_n=0 \ (n=3,\cdots D-1) $, we obtain the Lorentz covariant expression of the additional constraint (for $f_{\mu\nu}=\eta_{\mu\nu}$).
As we will see {in} the following section \ref{cons}, this is the most general case where the covariant expression of the constraint exists.

\subsection{$\beta_0 \neq 0,  \ \beta_1\neq0, \beta_n=0 \ (n=2,3,\cdots D-1) $ case}
\label{beta1}
Let us start with the case of $\beta_0 \neq 0,  \ \beta_1\neq0, \beta_n=0 \ (n=2,3,\cdots D-1) $ where the EoM (\ref{ly5}) is given by,
\begin{align}
E^{\mu\nu}= G^{\mu\nu} + \beta_0 g^{\mu\nu}+ \beta_1Y^{\mu\nu}_{(1)}(\ms) =0. \label{ooo4}
\end{align}
From the covariant constraint of the linearized model (\ref{bet3}), we can easily predict that the  following linear combination of $E^{\mu\nu}$ may be independent of any second order derivative terms $\partial_\mu \partial_\nu g_{\mu_1 \nu_1}$,
\begin{align}
\phi_1 \equiv \nabla_\nu \left( \ms^{-1\nu}_{~~~~~\rho} \nabla_\mu E^{\mu\rho}\right)
+ \frac{\beta_1}{D-2} g_{\mu\nu} E^{\mu\nu}. \label{qw1}
\end{align}
Indeed, by linearizing the above quantity (\ref{qw1}) and using the background equation $E^{\mu\nu}=0$, we obtain the quantity (\ref{bet3}).
In this part, we show that the quantity (\ref{qw1}) does not contain any second order derivative terms $\partial_\mu \partial_\nu g_{\mu_1 \nu_1}$ in the full-nonlinear level, i.e., we will see that the equation $\phi_1=0$ (under the EoM $E^{\mu\nu}=0$) can be regarded as a covariant constraint.  
Although we {suffer again from} the problem due to the square root matrix, we can avoid this problem by using a method analogous to the linear case.

In the case of  $\beta_0 \neq 0,  \ \beta_1\neq0, \beta_n=0 (n=2,3,\cdots D-1) $, the quantity $\ms^{-1\nu}_{~~~~~\rho} \nabla_\mu E^{\mu\rho}$ in (\ref{qw1}) is given by,
\begin{align}
\frac{1}{\beta_1}\ms^{-1\nu}_{~~~~~\rho} \nabla_\mu E^{\mu\rho}
=\mathcal{S}^{-1\nu}_{~~~~~\rho} \nabla_\mu Y_{(1)}^{\mu \rho} 
=\mathcal{S}^{-1\mu\nu}\nabla_\mu  \mathcal{S} - \mathcal{S}^{-1\nu}_{~~~~~\rho} 
\nabla_\mu \mathcal{S}^{\mu \rho} . \label{co4}
\end{align}
{The definitions of the covariant derivative of the ``matrices" $S_{\mu_1 \nu_1}$ and $f_{\mu_1 \nu_1}$ are not changed from the definition of a usual second rank tensor, i.e.,}
\begin{align}
 &\nabla_\mu \mathcal{S}_{\mu_1 \nu_1} \equiv \partial_\mu \mathcal{S}_{\mu_1 \nu_1}
- \Gamma_{\mu \mu_1}^{\rho} \mathcal{S}_{\rho\nu_1}
-\Gamma_{\mu \nu_1}^{\rho} \mathcal{S}_{\mu_1 \rho}, \notag \\
&\nabla_\mu f_{\mu_1 \nu_1} \equiv \partial_\mu f_{\mu_1 \nu_1}
- \Gamma_{\mu \mu_1}^{\rho} f_{\rho\nu_1}
-\Gamma_{\mu \nu_1}^{\rho} f_{\mu_1 \rho}.
\label{co3}
\end{align}
{We should note that the covariant derivative of the square root matrix, $\nabla_\mu \ms_{\mu_1 \nu_1}$, depends on the first derivative $\partial_\lambda g_{\mu\nu}$ through not only the Levi-Civita connection but also the partial derivative of the square root matrix, $\partial_\mu \ms_{\mu_1\nu_1}$.
If we substitute the expression (\ref{co4}) into the quantity (\ref{qw1}) straightforwardly, we find that there is the second order covariant derivative $\nabla_\mu \nabla_\nu \ms_{\mu_1 \nu_1}$ which contains second order partial derivative $\partial_\mu \partial_\nu \ms_{\mu_1 \nu_1}$. 
However, the dependence of $\partial_\mu \partial_\nu \ms_{\mu_1 \nu_1}$ on $\partial_\mu \partial_\nu g_{\mu_1 \nu_1}$ is not obvious and we cannot identify the dependence of (\ref{qw1}) on  $\partial_\mu \partial_\nu g_{\mu_1 \nu_1}$ in this way.
On the other hand, the covariant derivative of the fiducial metric, $\nabla_\mu f_{\mu_1\nu_1}$, depends on the first derivative $\partial_\lambda g_{\mu\nu}$ only through the Levi-Civita connection because the partial derivative $\partial_\mu f_{\mu_1 \nu_1}$ is independent of $\partial_\mu g_{\mu_1 \nu_1}$.
Hence, if we can rewrite $\nabla_\mu \nabla_\nu \ms_{\mu_1 \nu_1}$ in terms of $\nabla_\mu \nabla_\nu f_{\mu_1 \nu_1}$, we can identify the dependence of (\ref{qw1}) on $\partial_\mu \partial_\nu g_{\mu_1 \nu_1}$.}
Then, we try to express the quantity $\nabla_\nu \left(\mathcal{S}^{-1\nu\sigma} \nabla_\mu E^{\mu}_{~\sigma} \right)$ in (\ref{qw1}) so that the second order covariant derivatives are only acting on the fiducial metric $f_{\mu\nu}$.

Let us eliminate the covariant derivative from the square root matrix in Eq.(\ref{co4}).
From the relation, 
\begin{align}
\nabla_\lambda \ms^{\mu}_{~~\nu} \ms^{\nu}_{~~\rho} + \ms^{\mu}_{~~\nu} \nabla_\lambda \ms^{\nu}_{~~\rho}
= g^{\mu\nu} \nabla_\lambda f_{\nu \rho}, \label{kk1}
\end{align}
which can be derived by taking the covariant derivative of the relation (\ref{ly1}), 
we can show the following relation{s},
\begin{align}
&\nabla_\mu \mathcal{S} = 
\frac{1}{2} \mathcal{S}^{-1\alpha \beta} \nabla_\mu f_{\alpha \beta}, \label{cco3} \\
&2\mathcal{S}^{-1 (\nu}_{~~~~~\rho} \nabla_\mu \mathcal{S}^{\mu) \rho}
= S^{-1\mu \rho} \left( \nabla_\mu f_{\rho \sigma} \right) \mathcal{S}^{-1 \sigma \nu}.\label{cco1}
\end{align}
We find that the quantity $\nabla_\mu \ms$ in (\ref{co4}) can perfectly be rewritten in terms of $\nabla_\mu f_{\alpha \beta}$ by using (\ref{cco3}). On the other hand, from (\ref{cco1}),  we can determine the symmetric part $\mathcal{S}^{-1(\nu}_{~~~~~\rho} \nabla_\mu \mathcal{S}^{\mu) \rho}$ of $\mathcal{S}^{-1\nu}_{~~~~~\rho} \nabla_\mu \mathcal{S}^{\mu \rho}$ in (\ref{co4}). 

%with respect to $\mu\nu$.
Hence, we obtain the following expression, 
\begin{align}
\mathcal{S}^{-1\nu}_{~~~~~\rho} \nabla_\mu Y_{(1)}^{\mu \rho} (\ms)
= \frac12 \left(\mathcal{S}^{-1\mu\nu} \mathcal{S}^{-1\mu_1 \nu_1} 
-\mathcal{S}^{-1\mu (\mu_1}  \mathcal{S}^{-1 \nu_1) \nu} \right) \nabla_\mu f_{\mu_1 \nu_1}
-\mathcal{S}^{-1[\nu}_{~~~~~\rho} \nabla_\mu \mathcal{S}^{\mu] \rho}. \label{dy1}
\end{align}
Although there remains a covariant derivative acting on $\mathcal{S}$ in the last term in (\ref{dy1}), we should note that this covariant derivative turn into curvatures by taking an additional divergence because of the anti-symmetry of the superscripts $\mu\nu$.
Hence, we obtain,
\begin{align}
\nabla_\nu\left(\mathcal{S}^{-1\nu}_{~~~~~\rho} \nabla_\mu Y_{(1)}^{\mu \rho} (\ms)\right)=& \frac12  \left(\mathcal{S}^{-1\mu\nu} \mathcal{S}^{-1\mu_1 \nu_1} 
-\mathcal{S}^{-1\mu (\mu_1}  \mathcal{S}^{-1 \nu_1) \nu} \right) \nabla_\nu \nabla_\mu f_{\mu_1 \nu_1} \notag \\
&
+\frac12 R -\frac12 R^{\mu_1 \mu_2 \nu_1 \nu_2} \mathcal{S}^{-1}_{\mu_1 \nu_1} 
\mathcal{S}_{\mu_2 \nu_2} + \Phi_{(1)}, \label{co5.5} \\
\Phi_{(1)} \equiv&  \frac12  \nabla_\nu\left(\mathcal{S}^{-1\mu\nu} \mathcal{S}^{-1\mu_1 \nu_1} 
-\mathcal{S}^{-1\mu (\mu_1}  \mathcal{S}^{-1 \nu_1) \nu} \right) \cdot \nabla_\mu f_{\mu_1 \nu_1}
-\nabla_\nu \mathcal{S}^{-1[\nu}_{~~~~~\rho} 
 \cdot \nabla_\mu \mathcal{S}^{\mu] \rho} .\label{co6}
\end{align}
Here, $\Phi_{(1)}$ is the term which does not depend on any second order derivatives $\partial_\mu \partial_\nu g_{\alpha \beta}$. The other terms in (\ref{co5.5}) depend on second order derivatives $\partial_\mu \partial_\nu g_{\alpha \beta}$. The terms with curvatures, which are the second and the third terms in (\ref{co5.5}), are contributions from the anti-symmetric part $\mathcal{S}^{-1[\nu}_{~~~~~\rho} 
\nabla_{[\nu} \nabla_{\mu]} \mathcal{S}^{\mu] \rho}$. 

{We find that the only term depending on the second order covariant derivatives, the first term in (\ref{co5.5}), is expressed in terms of $\nabla_\nu \nabla_\mu f_{\mu_1 \nu_1}$. 
Because $\partial_\mu \partial_\nu f_{\mu_1 \nu_1}$ in $\nabla_\nu \nabla_\mu f_{\mu_1 \nu_1}$  does not depend on $\partial_\mu \partial_\nu g_{\alpha \beta}$,}
we can pick up the terms depending on the second order derivative $\partial_\mu \partial_\nu g_{\alpha \beta}$ from Eq.(\ref{co5.5}).
The parts depending on the second order derivative $\partial_\mu \partial_\nu g_{\alpha \beta}$ are given by,
\begin{align}
\left(\mathcal{S}^{-1\mu\nu} \mathcal{S}^{-1\mu_1 \nu_1} 
-\mathcal{S}^{-1\mu (\mu_1}  \mathcal{S}^{-1 \nu_1) \nu} \right) 
\nabla_\mu \nabla_\nu f_{\mu_1 \nu_1} &\supset 
-2\left(\mathcal{S}^{-1\mu\nu} \mathcal{S}^{-1\mu_1 \nu_1} 
-\mathcal{S}^{-1\mu (\mu_1}  \mathcal{S}^{-1 \nu_1) \nu} \right) 
 \partial_\mu \Gamma_{\rho,\nu (\mu_1} f_{\nu_1) }^{~~~\rho} \notag \\
&= \frac12 \mathcal{S}^{-1 \mu_1 \nu_1} \mathcal{S}^{\mu_2 \nu_2} \left[ 2\partial_{\mu_1} \partial_{\mu_2} g_{\nu_1 \nu_2} - \partial_{\mu_1} \partial_{\nu_1} g_{\mu_2 \nu_2} -\partial_{\mu_2} \partial_{\nu_2} g_{\mu_1 \nu_1} \right] \label{co8}
, \\
 \mathcal{S}^{-1 \mu_1 \nu_1} \mathcal{S}^{\mu_2 \nu_2} R_{\mu_1 \mu_2 \nu_1 \nu_2}
&\supset 2 \mathcal{S}^{-1 \mu_1 \nu_1} \mathcal{S}^{\mu_2 \nu_2}
\partial_{[\mu_1|} \Gamma_{\nu_1,|\mu_2] \nu_2} \notag \\
&= \frac12 \mathcal{S}^{-1 \mu_1 \nu_1} \mathcal{S}^{\mu_2 \nu_2} \left[ 2\partial_{\mu_1} \partial_{\mu_2} g_{\nu_1 \nu_2} - \partial_{\mu_1} \partial_{\nu_1} g_{\mu_2 \nu_2} -\partial_{\mu_2} \partial_{\nu_2} g_{\mu_1 \nu_1} \right].\label{cco6}
\end{align}
Here, we define the lowered connection $\Gamma_{\rho,\mu\nu} \equiv g_{\rho\sigma} \Gamma^{\sigma}_{\mu\nu}$.
We find that the r.h.s of (\ref{co8}) and the r.h.s of (\ref{cco6}) are canceled with each other in Eq.(\ref{co5.5}).
Hence, the following quantity $\Psi_{(1)}$ does not depend on $\partial_\mu \partial_\nu g_{\alpha \beta}$,     
\begin{align}
\Psi_{(1)} \equiv \frac12\left(\mathcal{S}^{-1\mu\nu} \mathcal{S}^{-1\mu_1 \nu_1} 
-\mathcal{S}^{-1\mu (\mu_1}  \mathcal{S}^{-1 \nu_1) \nu} \right) \nabla_\nu \nabla_\mu f_{\mu_1 \nu_1}
 -\frac12 R^{\mu_1 \mu_2 \nu_1 \nu_2} \mathcal{S}^{-1}_{\mu_1 \nu_1} 
\mathcal{S}_{\mu_2 \nu_2}. \label{co7}
\end{align}
Finally, by substituting (\ref{co7}) into (\ref{co5.5}), we obtain,
\begin{align}
\nabla_\nu\left(\mathcal{S}^{-1\nu}_{~~~~~\rho} \nabla_\mu Y_{(1)}^{\mu \rho}(\ms) \right)=
\frac{R}{2} + \Phi_{(1)} + \Psi_{(1)} .  \label{iy1}
\end{align}
Because the quantities $\Phi_{(1)},\ \Psi_{(1)}$ are independent of $\partial_\mu \partial_\nu g_{\alpha \beta}$, the only term depending on $\partial_\mu \partial_\nu g_{\alpha \beta}$ is $R/2$.
It is obvious that $R/2$ can be canceled by the trace $E^\mu_{~\mu}$, and we find that the following quantity does not depend on $\partial_\mu \partial_\nu g_{\mu_1 \nu_1}$,
\begin{align}
\frac{\phi_1}{\beta_1} \equiv \frac{1}{\beta_1}\nabla_\nu\left(\mathcal{S}^{-1\nu}_{~~~~~\rho} \nabla_\mu E^{\mu \rho} \right) + \frac{1}{D-2} g_{\mu\nu} E^{\mu\nu}= 
\Phi_{(1)} + \Psi_{(1)} + \beta_0\frac{D}{D-2} + \beta_1 \frac{D-1}{D-2} \ms, \label{by3}
\end{align}
which can be derived by using (\ref{iy1}) and (\ref{ooo4}).
In the case of $f_{\mu\nu} =\eta_{\mu\nu}$, 
this linear combination can be regarded as a Lorentz scalar function.
Therefore, in the case of $\beta_0 \neq 0,  \ \beta_1\neq0, \beta_n=0 \ (n=2,3,\cdots D-1) $, the equation $\phi_1=0$ (under the EoM $E^{\mu\nu}=0$) can be regarded as a covariant constraint.

\subsection{$\beta_0 \neq 0,  \ \beta_1\neq0, \ \beta_2\neq0 \ , \beta_n=0 \ (n=3,\cdots D-1) $ case}
Next, we derive the covariant constraint for the case of $\beta_0 \neq 0,  \ \beta_1\neq0, \ \beta_2\neq0 \ , \beta_n=0 \ (n=3,\cdots D-1) $ where the EoM (\ref{ly5}) is given by,
\begin{align}
E^{\mu\nu}= G^{\mu\nu} + \beta_0 g^{\mu\nu}+ \beta_1Y^{\mu\nu}_{(1)}(\ms)
+ \frac{\beta_2}{2} Y^{\mu\nu}_{(2)}(\ms) =0.\label{ooo2}
\end{align}
This is the most general case where the covariant expression of the additional constraint is possible, because, in the following section \ref{cons}, we will see that there are no covariant form of the additional constraint in other cases.
%The covariant constraint derived in this part is the most general form, because, in the following section \ref{cons}, we will see that there are no covariant form of the additional constraint in other cases.

Let us calculate the following quantity, 
\begin{align}
\nabla_\nu \left( \ms^{-1\nu}_{~~~~~\rho} \nabla_\mu E^{\mu \rho}\right)
= \beta_1 \nabla_\nu \left( \ms^{-1\nu}_{~~~~~\rho} \nabla_\mu Y^{\mu\rho}_{(1)}(\ms) \right)  + \beta_2\nabla_\nu \left( \ms^{-1\nu}_{~~~~~\rho} \nabla_\mu Y^{\mu\rho}_{(2)} (\ms) \right). \label{bb1}
\end{align}
as we have done in the previous part \ref{beta1}.
The $\beta_1$-term $ \nabla_\nu \left( \ms^{-1\nu}_{~~~~~\rho} \nabla_\mu Y^{\mu\rho}_{(1)}(\ms) \right)$ in the above equation (\ref{bb1}) have been calculated in (\ref{iy1}) of the previous part \ref{beta1}. 
Then, we now rewrite the $\beta_2$-term $ \nabla_\nu \left( \ms^{-1\nu}_{~~~~~\rho} \nabla_\mu Y^{\mu\rho}_{(2)}(\ms) \right)$ so that the second order covariant derivatives are only acting on $f_{\mu\nu}$.
By using the expansion formula given in (\ref{ky1}),
\begin{align}
Y^{\mu\nu}_{(n)} (\ms)= \sum_{k=0}^{n} (-1)^k e_{(n-k)} (\ms ) \left[\ms^k \right]^{\mu\nu}, 
\end{align}
$\ms^{-1\nu}_{~~~~~\rho}\nabla_\mu Y^{\mu\rho}_{(2)}(\ms)$ can be expressed as,
\begin{align}
\ms^{-1\nu}_{~~~~~\rho}\nabla_\mu Y^{\mu\rho}_{(2)}(\ms)
= \ms^{-1\nu}_{~~~~~\rho}\nabla^\rho e_{(2)} (\ms)- g^{\mu\nu} \nabla_\mu \ms -\ms \ms^{-1\nu}_{~~~~~\rho} \nabla_\mu \ms^{\mu\rho}
+\ms^{-1\nu}_{~~~~~\rho} \nabla_\mu\left[ \ms^2 \right]^{\mu\rho}. \label{cco2}
\end{align}
We find that the second and the third terms of the right hand side of the above equation (\ref{cco2}) can be rewritten by using Eqs.(\ref{cco3}) and (\ref{cco1}) given in the previous part \ref{beta1}.
Furthermore, the forth term of the right hand side of Eq.(\ref{cco2}) is trivially rewritten as $\ms^{-1\nu}_{~~~~~\rho} \nabla_\mu f^{\mu\rho}$. On the other hand, the first term of the right hand side of Eq.(\ref{cco2}) can be rewritten by using the following relation,
\begin{align}
\nabla_\mu e_{(n+1)}(\ms) = \frac{1}{2}Y^{\rho \sigma}_{(n)}(\ms) \nabla_\mu f_{\rho \lambda}  \ms^{-1\lambda}_{~~~~~\sigma}, \label{ky2}
\end{align}
which can be derived by using the identity obtained by multiplying $\left[\ms^{-1} Y_{(n)}(\ms)\right]_{\mu}^{~~\rho}$ on the both sides of the relation (\ref{kk1}) and use the commutativity $[Y_{(n)}(\ms),\ms]^{\mu\nu}=0$, as we have shown Eq.(\ref{oy2}).
Hence, by using the identities (\ref{cco3}), (\ref{cco1}), and (\ref{ky2}), we obtain,
\begin{align}
&\ms^{-1\nu}_{~~~~~\rho} \nabla_\mu  Y^{\mu\rho}_{(2)}(\ms)
= \frac{1}{2} \left[\ms \ms^{-1\mu\nu(\mu_1 \nu_1)} 
- \ms^{-1\mu\nu} g^{\mu_1 \nu_1} -g^{\mu\nu} \ms^{-1\mu_1 \nu_1} 
+ 2 \ms^{-1 \nu(\mu_1} g^{\nu_1) \mu}  \right]\nabla_\mu f_{\mu_1 \nu_1}
- \ms \ms^{-1[\nu}_{~~\rho} \nabla_\mu \ms ^{\mu] \rho}, \notag \\
&\ms^{-1\mu\nu\mu_1 \nu_1} \equiv \mathcal{S}^{-1\mu\nu} \mathcal{S}^{-1\mu_1 \nu_1} 
-\mathcal{S}^{-1\mu \nu_1}  \mathcal{S}^{-1 \mu_1 \nu}. \label{iii1}
\end{align}
The divergence of the above relation (\ref{iii1}) is given by,
\begin{align} 
&\nabla_\nu \left( \ms^{-1\nu}_{~~~~~\rho} \nabla_\mu  Y^{\mu\rho}_{(2)}(\ms)\right)
=\frac{\ms}{2} R+\frac12\left[- \ms^{-1\mu\nu} g^{\mu_1 \nu_1} -g^{\mu\nu} \ms^{-1\mu_1 \nu_1} 
+ 2 \ms^{-1 \nu(\mu_1} g^{\nu_1) \mu}  \right]\nabla_\nu  \nabla_\mu f_{\mu_1 \nu_1}
 +\ms \Psi_{(1)}+\Phi_{(2)}, \label{iy2.5} \\ 
&\Phi_{(2)} \equiv \frac{1}{2} \nabla_\nu\left[\ms \ms^{-1\mu\nu(\mu_1 \nu_1)} 
- \ms^{-1\mu\nu} g^{\mu_1 \nu_1} -g^{\mu\nu} \ms^{-1\mu_1 \nu_1} 
+ 2 \ms^{-1 \nu(\mu_1} g^{\nu_1) \mu}  \right] \cdot \nabla_\mu f_{\mu_1 \nu_1}
- \nabla_\nu \left( \ms \ms^{-1[\nu}_{~~\rho} \right) \cdot \nabla_\mu \ms ^{\mu] \rho}.\label{iy3}
\end{align}
Here, we define the $\Phi_{(2)}$ which does not depend on $\partial_\mu \partial_\nu g_{\alpha \beta}$.
Furthermore, we express the equation (\ref{iy2.5}) by using the quantity $\Psi_{(1)}$ defined in (\ref{co7}).
We have succeeded to express the second order covariant derivative terms in the quantity $\nabla_{\nu}\left(\ms^{-1\nu}_{~~~~~\rho} \nabla_\mu  Y^{\mu\rho}_{(2)}(\ms)\right)$ by using $\nabla_\mu \nabla_\nu f_{\mu_1 \nu_1}$.

As we have done in Eqs.(\ref{co8}) and (\ref{cco6}), by decomposing $R$ and $ \ \nabla_\mu \nabla_\nu f_{\mu_1 \nu_1}$ in Eq.(\ref{iy2.5}) into the terms depending on $\partial_\mu \partial_\nu g_{\alpha \beta}$ and the other terms, the dependence of  $\nabla_\nu \left( \ms^{-1\nu}_{~~~~~\rho} \nabla_\mu  Y^{\mu\rho}_{(2)}(\ms)\right)$ on $\partial_\mu \partial_\nu g_{\alpha \beta}$ reads,
\begin{align}
\nabla_\nu \left( \ms^{-1\nu}_{~~~~~\rho} \nabla_\mu  Y^{\mu\rho}_{(2)}(\ms)\right) &\supset \frac{\ms}{2} R+\frac12\left[- \ms^{-1\mu\nu} g^{\mu_1 \nu_1} -g^{\mu\nu} \ms^{-1\mu_1 \nu_1} 
+ 2 \ms^{-1 \nu(\mu_1} g^{\nu_1) \mu}  \right]\nabla_\nu  \nabla_\mu f_{\mu_1 \nu_1} \notag \\
&\supset -\frac12 g^{\mu\nu\mu_1 \nu_1 \mu_2 \nu_2}  \partial_\mu \partial_\nu g_{\mu_1 \nu_1} \ms_{\mu_2 \nu_2}. \label{oy3}
\end{align}
On the other hand, the dependence of $\ms_{\mu\nu}E^{\mu\nu}$ on $\partial_\mu \partial_\nu g_{\alpha \beta}$ reads,
\begin{align}
\ms_{\mu\nu}E^{\mu\nu} &\supset \ms_{\mu\nu}G^{\mu\nu} \notag \\
&\supset \frac12 g^{\mu\nu\mu_1 \nu_1 \mu_2 \nu_2}  \partial_\mu \partial_\nu g_{\mu_1 \nu_1} \ms_{\mu_2 \nu_2}. \label{ooo1}
\end{align}
Therefore, from the relations (\ref{iy1}), (\ref{ooo2}), (\ref{bb1}), (\ref{oy3}) and (\ref{ooo1}), we find that the following linear combination of $E^{\mu\nu}$ does not depend on $\partial_\mu \partial_\nu g_{\alpha \beta}$,
\begin{align}
\phi_{cov}\equiv& \nabla_\nu \left( \ms^{-1\nu}_{~~~~~\rho} \nabla_\mu E^{\mu\rho}\right)
+ \frac{\beta_1}{D-2} g_{\mu\nu} E^{\mu\nu} + \beta_2 \ms_{\mu\nu}  E^{\mu\nu}
\notag \\
=& \beta_1 \left( \Phi_{(1)} + \Psi_{(1)} + \sum_{n=0}^2 \beta_n\frac{D-n}{D-2} e_{(n)}(\ms) \right)\notag \\
&+ \beta_2 \left( \ms \Psi_{(1)}+ \Phi_{(2)} +\Psi_{(2)}  + \sum_{n=0}^{2}\beta_{n} (n+1) e_{(n+1)} (\ms) \right), \notag \\
\Psi_{(2)} &\equiv \frac{\ms}{2} R+\frac12\left[- \ms^{-1\mu\nu} g^{\mu_1 \nu_1} -g^{\mu\nu} \ms^{-1\mu_1 \nu_1} 
+ 2 \ms^{-1 \nu(\mu_1} g^{\nu_1) \mu}  \right]\nabla_\nu  \nabla_\mu f_{\mu_1 \nu_1}
+\ms_{\mu\nu}G^{\mu\nu} \notag \\
&=\frac12\left[- \ms^{-1\mu\nu} g^{\mu_1 \nu_1} -g^{\mu\nu} \ms^{-1\mu_1 \nu_1} 
+ 2 \ms^{-1 \nu(\mu_1} g^{\nu_1) \mu}  \right]\nabla_\nu  \nabla_\mu f_{\mu_1 \nu_1}
+\ms_{\mu\nu}R^{\mu\nu}. \label{covariant}
\end{align}
Here, the quantity $\Psi_{(2)}$ does not depend on $\partial_\mu \partial_\nu g_{\alpha \beta}$ because of Eqs.(\ref{oy3}) {and} (\ref{ooo1}).
Therefore, in the case of $\beta_0 \neq 0,  \ \beta_1\neq0, \ \beta_2\neq0 \ , \beta_n=0 \ (n=3,\cdots D-1) $, there is a covariant constraint $\phi_{cov}=0$.
As we will see in the following section \ref{cons}, 
this is the most general case where the covariant expression is possible.

%this is the most general form of the covariant constraint.

\section{Constraint structure}
\label{cons}
From this section, we extend the argument in the previous section \ref{cov} to the general case ($\beta_{n}\neq0,~ n=0,1,\cdots D-1$). Although the covariant expression is impossible in the general case, we can show the existence of the additional constraint.

For the first step, in this section,
the existence of the additional constraint can be shown by using the following identity,
\begin{align}
\nabla_\nu \left( \ms^{-1\nu}_{~~~~~\rho} \nabla_\mu Y^{\mu\rho}_{(n)}(\ms)\right)
=&-\frac12 Y^{\mu\nu\mu_1 \nu_1}_{(n-1)} (\ms)
\partial_\mu \partial_\nu g_{\mu_1 \nu_1} +(\text{terms without }\partial_\mu \partial_\nu g_{\mu_1 \nu_1}), \notag \\
 Y^{\mu\nu\mu_1 \nu_1}_{(n-1)} (\ms) \equiv&
\frac{1}{(n-1)!}  g^{\mu\nu\mu_1\nu_1\mu_2 \nu_2 \mu_3 \nu_3 \cdots \mu_n \nu_n}
\ms_{\mu_2 \nu_2} \ms_{\mu_3 \nu_3} \cdots \ms_{\mu_n \nu_n}. \label{ly8}
\end{align}
Although the general proof of this identity will be given in the following section \ref{pro}, we can confirm that the identity (\ref{ly8}) is valid for the case of $n=1, \ 2$ from the results (\ref{iy1}) {and} (\ref{oy3}) in the previous section \ref{cov}.
By using the identity (\ref{ly8}), we can easily show the existence of the additional constraint.

\subsection{$\beta_0 \neq 0,  \ \beta_1\neq0, \ \beta_2\neq0 \ , \beta_n=0 \ (n=3,\cdots D-1) $ case}
First, we would like to show again that the additional constraint takes the covariant form in the specific case $\beta_0 \neq 0,  \ \beta_1\neq0, \ \beta_2\neq0 \ , \beta_n=0 \ (n=3,\cdots D-1)$ which have been considered in the previous section \ref{cov}.
In this case, by using the identity (\ref{ly8}), the linear combination $\nabla_\nu \left( \ms^{-1\nu}_{~~~~~\rho} \nabla_\mu E^{\mu\rho}\right)$ becomes,
\begin{align}
\nabla_\nu \left( \ms^{-1\nu}_{~~~~~\rho} \nabla_\mu E^{\mu\rho}\right)
= -\frac{\beta_1}{2} g^{\mu\nu\mu_1 \nu_1} \partial_\mu  \partial_\nu g_{\mu_1 \nu_1}
-\frac{\beta_2}{2} g^{\mu\nu\mu_1\nu_1\mu_2 \nu_2}\partial_\mu  \partial_\nu g_{\mu_1 \nu_1}
\ms_{\mu_2 \nu_2}+(\text{terms without }\partial_\mu  \partial_\nu g_{\mu_1 \nu_1}).
\label{11/29/3}
\end{align}
On the other hand, the parts depending on the second order derivative $\partial_\mu  \partial_\nu g_{\mu_1 \nu_1}$ in the Einstein tensor $G^{\mu\nu}$ are given by,
\begin{align}
G^{\mu\nu} = \frac12 g^{\mu\nu\mu_1 \nu_1 \mu_2 \nu_2} \partial_{\mu_1}\partial_{\nu_1}
g_{\mu_2 \nu_2} + (\text{terms without }\partial_\mu  \partial_\nu g_{\mu_1 \nu_1}). \label{ly6}
\end{align}
Hence, the dependences of $g_{\mu\nu} E^{\mu\nu}, \ \ms_{\mu\nu}  E^{\mu\nu}$ on $\partial_\mu  \partial_\nu g_{\mu_1 \nu_1}$ are given by,
\begin{align}
&g_{\mu\nu} E^{\mu\nu} = \frac{D-2}{2}  g^{\mu_1 \nu_1 \mu_2 \nu_2} \partial_{\mu_1}\partial_{\nu_1}g_{\mu_2 \nu_2} + (\text{terms without }\partial_\mu  \partial_\nu g_{\mu_1 \nu_1}), \label{11/29/1} \\
&\ms_{\mu\nu}  E^{\mu\nu}= \frac12 g^{\mu_1 \nu_1 \mu_2 \nu_2\mu\nu}  \partial_{\mu_1}\partial_{\nu_1}
g_{\mu_2 \nu_2} \ms_{\mu\nu}
+ (\text{terms without }\partial_\mu  \partial_\nu g_{\mu_1 \nu_1})
\label{11/29/2}.
\end{align}
Here, we used the formula (\ref{Ap2}) and the symmetry (\ref{Appp1}) to rewrite Eqs.(\ref{11/29/1}) and (\ref{11/29/2}) respectively. 
Thus, from Eqs.(\ref{11/29/3}), (\ref{11/29/1}) and (\ref{11/29/2}), we find that the following covariant linear combination does not contain any second derivative terms $\partial_\mu  \partial_\nu g_{\mu_1 \nu_1}$,
\begin{align}
\phi_{cov} \equiv \nabla_\nu \left( \ms^{-1\nu}_{~~~~~\rho} \nabla_\mu E^{\mu\rho}\right)
+ \frac{\beta_1}{D-2} g_{\mu\nu} E^{\mu\nu} + \beta_2 \ms_{\mu\nu}  E^{\mu\nu}.
\end{align}
Therefore, in the case of $\beta_k=0 (3\le k)$, there is the covariant constraint $\phi_{cov}=0$.

\subsection{General case}
Although the additional constraint cannot be expressed as a covariant form in the general case $\beta_k \neq 0 \ (3\le k)$, we can easily show the existence of the additional constraint.

Let us focus on the second order ``time" derivative terms of (\ref{ly8}),
\begin{align}
\nabla_\nu \left( \ms^{-1\nu}_{~~~~~\rho} \nabla_\mu Y^{\mu\rho}_{(n)}(\ms)\right)
=&-\frac12 \frac{1}{(n-1)!}  g^{00} \theta^{\mu_1\nu_1\mu_2 \nu_2 \mu_3 \nu_3 \cdots \mu_n \nu_n} \ms_{\mu_2 \nu_2} \ms_{\mu_3 \nu_3} \cdots \ms_{\mu_n \nu_n}
\partial_0 \partial_0 g_{\mu_1 \nu_1}\notag \\ 
&+(\text{terms without }\partial_0 \partial_0 g_{\mu_1 \nu_1}) , \notag \\
\theta^{\mu_1 \nu_1 \mu_2 \nu_2 \cdots \mu_n \nu_n}
\equiv& \delta^{\mu_1~~\mu_2~~ \cdots \mu_n }_{~~~\rho_1~~\rho_2 ~~\cdots \rho_n}
\theta^{\rho_1\nu_1} \theta^{\rho_2 \nu_2} \cdots \theta^{\rho_n \nu_n} , \notag \\
\theta^{\mu\nu} \equiv& g^{\mu \nu} - \frac{g^{0\mu} g^{0\nu}}{g^{00}}. \label{ly9}
\end{align}
Here, we use the identity (\ref{klkl1}), and $\theta^{\mu\nu}$ is the projection operator living in $D-1$ dimensional space orthogonal to the time direction $g^0_\mu$. We lower indexes of $\theta^{\mu\nu}$ by using $g_{\mu\nu}$ and raise indexes by using $g^{\mu\nu}$.
We find that the quantity $\nabla_\nu \left( \ms^{-1\nu}_{~~~~~\rho} \nabla_\mu Y^{\mu\rho}_{(n)}(\ms)\right)$ does not depend on $g^{0\mu}g^{0\nu}\partial_{0}\partial_{0}g_{\mu\nu}$ nor $g^{0\mu}\theta^{\rho\nu}\partial_{0}\partial_{0}g_{\mu\nu}$.
In other words, the quantity $\nabla_\nu \left( \ms^{-1\nu}_{~~~~~\rho} \nabla_\mu Y^{\mu\rho}_{(n)}(\ms)\right)$ depends only on  $\theta^\mu_{\alpha}\theta^{\nu}_\beta \partial_0 \partial_0 g_{\mu \nu}$.

On the other hand, from (\ref{ly6}), the second order ``time" derivative terms in EoM (\ref{ly5}) are expressed as,
\begin{align}
&E^{\mu\nu} = \frac12 g^{00} \theta^{\mu\nu \mu_1 \nu_1} \partial_0 \partial_0 g_{\mu_1 \nu_1} + (\text{terms without }\partial_0 \partial_0 g_{\mu_1 \nu_1})^{\mu\nu}=0. 
\label{ly7}
\end{align}
The EoM (\ref{ly7}) can be solved with respect to the $\theta_\mu^\alpha \theta_\nu^{\beta} \partial_0 \partial_0 g_{\alpha \beta}$. In order to do that, we introduce an ``inverse matrix" $\theta^{-1}_{\mu\nu,\mu_1 \nu_1} $ of the operator $\theta^{(\mu\nu) (\mu_1 \nu_1)}$ so that it satisfies the following condition,
\begin{align}
&\theta^{-1}_{\mu\nu,\mu_1 \nu_1} \theta^{(\mu_1\nu_1) (\mu_2 \nu_2)} := \theta_\mu^{(\mu_2} \theta_\nu^{\nu_2)}, \notag \\
&0:= \theta^{-1}_{\mu \nu,\mu_1 \nu_1} g^{\mu0}=\theta^{-1}_{\mu \nu,\mu_1 \nu_1} g^{\nu0}
=\theta^{-1}_{\mu \nu,\mu_1 \nu_1} g^{\mu_10}=\theta^{-1}_{\mu \nu,\mu_1 \nu_1} g^{\nu_10}, \notag \\
&\theta_{\mu\nu,\mu_1 \nu_1}^{-1} := \theta_{(\mu\nu),(\mu_1 \nu_1)}^{-1}.
\end{align}
These conditions are uniquely solved as,
\begin{align}
\theta^{-1}_{\mu\nu,\mu_1 \nu_1} = \frac{\theta_{\mu\nu} \theta_{\mu_1 \nu_1}}{D-2}
-\theta_{\mu(\mu_1} \theta_{\nu_1)\nu}. \label{iii5}
\end{align}
Using the above operator (\ref{iii5}), we obtain, 
\begin{align}
\theta^{-1}_{\mu\nu,\mu_1 \nu_1}E^{\mu_1 \nu_1} = \frac{g^{00}}{2}\theta^{\alpha}_{\mu} \theta^\beta_{\nu} \partial_0 \partial_0 g_{\alpha \beta} +
 (\text{terms without }\partial_0  \partial_0 g_{\mu_1 \nu_1}). \label{ly10}
\end{align}

From the identity (\ref{ly9}) and the above relation (\ref{ly10}), we find that the following linear combination of $E^{\mu\nu}$ does not depend on any second time derivative $\partial_0 \partial_0 g_{\mu\nu}$,
\begin{align}
\psi \equiv \nabla_\nu \left( \ms^{-1\nu}_{~~~~~\rho} \nabla_\mu E^{\mu\rho}\right)
+ \sum_{n=1}^{D-1}  \frac{\beta_n}{(n-1)!}   \theta^{\mu_1\nu_1\mu_2 \nu_2 \mu_3 \nu_3 \cdots \mu_n \nu_n} \ms_{\mu_2 \nu_2} \ms_{\mu_3 \nu_3} \cdots \ms_{\mu_n \nu_n}
\theta^{-1}_{\mu_1\nu_1,\rho \sigma}E^{\rho \sigma}.\label{sum2}
\end{align}
or,
\begin{align}
\psi' \equiv& \nabla_\nu \left( \ms^{-1\nu}_{~~~~~\rho} \nabla_\mu E^{\mu\rho}\right)
+ \frac{\beta_1}{D-2} g_{\mu\nu} E^{\mu\nu} + \beta_2 \ms_{\mu\nu}  E^{\mu\nu} \notag \\
&+  \sum_{n=3}^{D-1}  \frac{\beta_n}{(n-1)!}   \theta^{\mu_1\nu_1\mu_2 \nu_2 \mu_3 \nu_3 \cdots \mu_n \nu_n} \ms_{\mu_2 \nu_2} \ms_{\mu_3 \nu_3} \cdots \ms_{\mu_n \nu_n}
\theta^{-1}_{\mu_1\nu_1,\rho \sigma}E^{\rho \sigma}. \label{sum3}
\end{align}
Hence, there is the additional constraint $\psi=0$ or $\psi'=0$. In the case of $f_{\mu\nu}=\eta_{\mu\nu}$, this combination (\ref{sum2}) (or (\ref{sum3})) is not Lorentz invariant but invariant for spacial rotations.
{This result is consistent with known several results.
\cite{bernard2} has concluded that the covariant constraint in the linearized model with metric formulation for $D=4$ exists only in the case of $\beta_3=0$.
\cite{viel3} have argued the covariant constraints in the vielvein formulation for $D=4$ and concluded that the covariant constraint exists only in the case of $\beta_3=0$.
These results are consistent with our result that the covariant constraint exists only in the case of $\beta_0 \neq 0,  \ \beta_1\neq0, \ \beta_2\neq0 \ , \beta_n=0 \ (n=3,\cdots D-1)$.}

\section{Proof of identity}
\label{pro}
In the previous section \ref{cons}, we have seen that the identity (\ref{ly8}) plays a crucial role for the existence of the additional constraint.
%BD ghost-freeness of the dRGT model.
In this section, we complete our argument 
%the proof of the existence of the additional constraint 
by proving the identity (\ref{ly8}) for the general case.

Let us explain how to show the identity (\ref{ly8}).
By using the expansion formula,
\begin{align}
Y^{\mu\nu}_{(n)} (\ms)= \sum_{k=0}^{n} (-1)^k e_{(n-k)} (\ms ) \left[\ms^k \right]^{\mu\nu},  
\label{hy7}
\end{align}
whose derivation is given in (\ref{ky1}) of the Appendix \ref{ap1},
we can express the quantity $\ms^{-1\nu}_{~~~~~\rho} \nabla_\mu Y^{\mu\rho}_{(n)}(\ms)$ in the left hand side of the identity (\ref{ly8}) as follows,
\begin{align}
\ms^{-1\nu}_{~~~~~\rho} \nabla_\mu Y^{\mu\rho}_{(n)} (\ms)= \sum_{k=0}^{n} (-1)^k 
\left[ \nabla_\mu e_{(n-k)} (\ms ) \left[\ms^{k-1} \right]^{\mu\nu}  +  e_{(n-k)} (\ms )
\ms^{-1\nu}_{~~~~~\rho} \nabla_\mu  \left[\ms^{k} \right]^{\mu\rho}\right].\label{hy7}
\end{align}

As we have done in the Sec.\ref{cov}, in order to confirm the identity (\ref{ly8}), we try to rewrite $\nabla_\mu \nabla_\nu \ms_{\alpha \beta}$ in $\nabla_\nu \left(\mathcal{S}^{-1\nu\sigma} \nabla_\mu {Y_{(1)}}^{\mu}_{~\sigma} \right)$ in terms of $\nabla_\mu \nabla_\nu f_{\alpha \beta}$ by using the relation,
\begin{align}
\nabla_\lambda \ms^{\mu}_{~~\nu} \ms^{\nu}_{~~\rho} + \ms^{\mu}_{~~\nu} \nabla_\lambda \ms^{\nu}_{~~\rho}
= g^{\mu\nu} \nabla_\lambda f_{\nu \rho}. \label{iii4}
\end{align}
By multiplying the matrix $\ms^n \ (n \in \mathbb{Z})$ on the above equation (\ref{iii4}) from the left hand side, and multiplying the matrix $\ms^m g \ (m \in \mathbb{Z})$ from the right hand side, we obtain more general formula,
\begin{align}
[\ms^n \nabla_\lambda \ms \ms^{m+1}  + \ms^{n+1} \nabla_\lambda \ms \ms^m]^{\mu\nu}
= [\ms^{n} \nabla_\lambda f \ms^{m}]^{\mu\nu}. \label{hy3}
\end{align}
Here, 
we define the matrix multiplication $[AB]^{\mu\nu} \equiv A^\mu_{~~\rho} B^{\rho\nu}$, and 
we omit the metric $g$ in the definition of the matrix multiplication.
In particular, in the case of $n=m$, the left hand side of the above equation (\ref{hy3}) can be expressed as a symmetrization of a matrix product,
\begin{align}
[\ms^n \nabla_\lambda \ms \ms^{n+1}]^{(\mu\nu)}
= \frac12 [\ms^{n} \nabla_\lambda f \ms^{n}]^{\mu\nu}. \label{hy5}
\end{align}
 
As we have derived Eq.(\ref{ky2}), we obtain the following formula from Eq.(\ref{hy3}),
\begin{align}
\nabla_\lambda e_{(n)} (\ms) 
=\frac12   {Y_{(n)}}^{\mu\nu} (\ms) \nabla_\lambda f_{\mu \rho} \ms^{-1\rho}_{~~~~~\nu} . \label{hy4}
\end{align}
Furthermore, we can rewrite the quantity $\ms^{-1\nu}_{~~~~~\rho} \nabla_{\mu} \left[\ms^{2n+1} \right]^{\mu\rho} \ (n = 0,1,2,3 \cdots )$ from the relation (\ref{hy5}),
\begin{align}
\ms^{-1\nu}_{~~~~~\rho} \nabla_{\mu} \left[\ms^{4m+1} \right]^{\mu\rho}
&= \left[ \left( \nabla_{\mu}f^{m}\ms^{2m+1}+
\ms^{2m+1}\nabla_{\mu}f^{m} \right)\ms^{-1}
+ \frac12 \ms^{2m-1} \nabla_\mu f \ms^{2m-1} \right]^{\mu\nu}
+ \left[ \ms^{2m} \nabla_\mu \ms \ms^{2m-1} \right]^{[\mu\nu]}, \label{gy4} \\
\ms^{-1\nu}_{~~~~~\rho} \nabla_{\mu} \left[\ms^{4m+3} \right]^{\mu\rho}
&= \left[ \left( \nabla_{\mu}f^{m+1}\ms^{2m+1}+
\ms^{2m+1}\nabla_{\mu}f^{m+1} \right)\ms^{-1}
- \frac12 \ms^{2m} \nabla_\mu f \ms^{2m} \right]^{\mu\nu}
 -\left[ \ms^{2m+1} \nabla_\mu \ms \ms^{2m} \right]^{[\mu\nu]},\label{hy6}
\end{align}
whose derivations are given in Appendix \ref{ap2}.
Here, we separate the set of $n$ into the even case $n=2m \ (m=0,1,2,\cdots)$ and the odd case $n=2m+1 \ (m=0,1,2,3\cdots)$.
In the above equations (\ref{hy4}), (\ref{gy4}) and (\ref{hy6}), all the covariant derivative terms except for the anti-symmetric terms with respect to $\mu\nu$ in (\ref{gy4}) and (\ref{hy6}) are expressed in terms of $\nabla_\lambda f_{\mu\nu}$. This fact means that we can rewrite all the terms with $\partial_{\mu}\partial_{\nu}g_{\alpha\beta}$ in the quantity $\nabla_\nu \left(\ms^{-1\nu}_{~~~~~\rho} \nabla_\mu Y^{\mu\rho}_{(n)}(\ms)\right)$ as in terms of $\nabla_\mu \nabla_\nu f_{\alpha \beta}$ and Riemann curvatures because the covariant derivative of $\left[ \ms^{2m} \nabla_\mu \ms \ms^{2m-1} \right]^{[\mu\nu]}$ and $\left[ \ms^{2m+1} \nabla_\mu \ms \ms^{2m} \right]^{[\mu\nu]}$ in Eqs.(\ref{gy4}) and (\ref{hy6}) will turn into curvatures by taking an additional divergence.
Therefore, in principle, we can pick up the terms depending on $\partial_{\mu} \partial_\nu g_{\alpha \beta}$ by using the identities (\ref{hy4}), (\ref{gy4}) and (\ref{hy6}) from $\nabla_\nu \left(\ms^{-1\nu}_{~~~~~\rho} \nabla_\mu Y^{\mu\rho}_{(n)}(\ms)\right)$ in the identity (\ref{ly8}).

By substituting (\ref{hy4}) into the divergence of (\ref{hy7}), we obtain,
\begin{align}
\nabla_\mu \left(\ms^{-1\nu}_{~~~~~\rho} \nabla_\mu Y^{\mu \rho}_{(n)} (\ms) \right)
=\frac12 \sum_{K=0}^{n-1} (-1)^K
 e_{(n-K-1)} (\ms) T_{(K)} + (\text{terms without }\partial_\mu \partial_\nu g_{\alpha \beta} ).
\label{gy10}
\end{align}
Here, we define the following quantity,
\begin{align}
T_{(K)} \equiv \sum_{k=0}^{K} \left[ \ms^{k-1} \right]^{\mu\nu} \left[ \ms^{K-k-1} \right]^{\mu_1 \nu_1} \nabla_\nu \nabla_\mu f_{\mu_1\nu_1} 
-2 \nabla_\nu \left( \ms^{-1\nu}_{~~~~~\rho} \nabla_\mu \left[ \ms^{K+1} \right]^{\mu \rho}\right).
\label{gy1}
\end{align}
By considering the several cases $K=4m, 4m+1, 4m+2, 4m+3 \ (m=0,1,2,3\cdots)$ for $T_{(K)}$, we obtain,
\begin{align}
T_{(4m)} \sim& \left\{ \sum_{k=0}^{m-1} \tilde{T}^{\mu\nu\mu_1 \nu_1} (\ms^{2k-1},\ms^{4m-2k-1})
+ \sum_{k=0}^{m-1} \tilde{T}^{\mu\nu\mu_1 \nu_1}(\ms^{4m-2k-2},\ms^{2k})
+ \frac12 \tilde{T}^{\mu\nu\mu_1 \nu_1} (\ms^{2m-1},\ms^{2m-1}) \right\}
\nabla_\nu \nabla_\mu f_{\mu_1 \nu_1} \notag \\
&- 2\left[ \ms^{2m}\nabla_\nu \nabla_\mu \ms \ms^{2m-1}\right]^{[\mu\nu]}, \label{gy5} \\
T_{(4m+1)} \sim& \left\{ 
\sum_{k=0}^{m-1} \tilde{T}^{\mu\nu\mu_1 \nu_1} (\ms^{2k-1},\ms^{4m-2k}) +\sum_{k=0}^{m} \tilde{T}^{\mu\nu\mu_1\nu_1} (\ms^{4m-2k-1},\ms^{2k}) 
\right\} \nabla_\nu \nabla_\mu f_{\mu_1 \nu_1},  \label{gy6}\\
T_{(4m+2)} \sim& \left\{ \sum_{k=0}^{m-1} \tilde{T}^{\mu\nu\mu_1 \nu_1} (\ms^{4m-2k},\ms^{2k})
+\sum_{k=0}^m \tilde{T}^{\mu\nu\mu_1 \nu_1} (\ms^{2k-1},\ms^{4m-2k+1})
+ \frac12 \tilde{T}^{\mu\nu\mu_1 \nu_1}(\ms^{2m},\ms^{2m}) \right\}
\nabla_\nu \nabla_\mu f_{\mu_1 \nu_1} \notag \\
&+2 \left[ \ms^{2m+1} \nabla_\nu \nabla_\mu \ms \ms^{2m}\right]^{[\mu\nu]} , \label{gy7} \\
T_{(4m+3)}
\sim &\left\{ 
\sum_{k=0}^{m} \tilde{T}^{\mu\nu\mu_1 \nu_1} (\ms^{2k-1}, \ms^{4m-2k+2})
+ \sum_{k=0}^{m}\tilde{T}^{\mu\nu\mu_1 \nu_1} (\ms^{4m-2k+1}, \ms^{2k})\right\} \nabla_\nu \nabla_\mu f_{\mu_1 \nu_1}, \label{gy8}
\end{align}
whose derivation is summarized in Appendix \ref{ap2}.
Here, we use Eqs.(\ref{gy4}) {and} (\ref{hy6}) for calculating Eqs.(\ref{gy5}) {and} (\ref{gy7}), respectively. 
Furthermore, we define the {notation} ``$\sim$" which means the equivalence up to terms independent of $\partial_\mu \partial_\nu g_{\alpha \beta}$, and define the bi-linear function,
\begin{align}
\tilde{T}^{\mu\nu\mu_1 \nu_1}(\ms^{a},\ms^{b})
 \equiv  \left[ \ms^{a} \right]^{\mu\nu} \left[ \ms^{b} \right]^{\mu_1\nu_1} + \left[ \ms^{a} \right]^{\mu_1\nu_1} \left[ \ms^{b} \right]^{\mu\nu} -2  \left[ \ms^{b} \right]^{\mu(\mu_1} \left[ \ms^{a} \right]^{\nu_1)\nu},
\end{align}
for any $a,b\in \mathbb{Z}$.
We should note that this tensor is {\it not} symmetric for the permutation $\mu\leftrightarrow \nu$, but satisfies the following relation,
\begin{align}
\tilde{T}^{\mu\nu\mu_1 \nu_1} (\ms^{a},\ms^{b}) =
\tilde{T}^{\nu\mu\mu_1 \nu_1} (\ms^{b},\ms^{a}).
\end{align}
Then, in the case of $a=b$, this tensor becomes symmetric with respect to $\mu\nu$, $\tilde{T}^{\mu\nu\mu_1 \nu_1}(\ms^a,\ms^a) = \tilde{T}^{(\mu\nu)\mu_1 \nu_1}(\ms^a,\ms^a)$.

Next, we identify the terms depending on $\partial_\mu \partial_\nu g_{\alpha \beta}$ in Eqs(\ref{gy5}), (\ref{gy6}), (\ref{gy7}) and (\ref{gy8}). We should note that all the terms in Eqs.(\ref{gy5}), (\ref{gy6}), (\ref{gy7}) and (\ref{gy8}) take the form as $\tilde{T}^{\mu\nu\mu_1 \nu_1} (\ms^a, \ms^b)\nabla_\nu \nabla_\mu f_{\mu_1 \nu_1}$ or $\left[ \ms^{a} \nabla_\nu \nabla_\mu \ms \ms^{b}\right]^{[\mu\nu]}$. 
Thus, we can identify these dependences on $\partial_\mu \partial_\nu g_{\alpha \beta}$ by using the following relations,
\begin{align}
&\tilde{T}^{\mu\nu\mu_1 \nu_1}(\ms^n,\ms^m)
\nabla_\nu \nabla_\mu f_{\mu_1 \nu_1} = - T^{\mu\nu\mu_1 \nu_1} (\ms^{n+2},\ms^m) \partial_\mu \partial_\nu g_{\mu_1 \nu_1} 
+(\text{terms without $\partial_\mu \partial_\nu g_{\alpha \beta}$}) , \notag \\
&\left[\ms^n \right]^{\mu_1 \nu_1} \left[ \ms^m\right]^{\mu_2 \nu_2} R_{\mu\mu_1 \nu\nu_1} 
= -\frac12 T^{\mu\nu\mu_1 \nu_1} (\ms^n,\ms^m) \partial_\mu \partial_\nu g_{\mu_1 \nu_1} 
+(\text{terms without $\partial_\mu \partial_\nu g_{\alpha \beta}$}), \label{gy9}
\end{align}
which can be shown by straightforward calculation.
Here, we define the new bilinear function,
\begin{align}
T^{\mu\nu\mu_1 \nu_1} (\ms^a,\ms^b) \equiv  \left[ \ms^{a} \right]^{\mu\nu} \left[ \ms^{b} \right]^{\mu_1\nu_1} + \left[ \ms^{a} \right]^{\mu_1\nu_1} \left[ \ms^{b} \right]^{\mu\nu} -  \left[ \ms^{b} \right]^{\mu(\mu_1} \left[ \ms^{a} \right]^{\nu_1)\nu}
-\left[ \ms^{b} \right]^{\nu(\mu_1} \left[ \ms^{a} \right]^{\nu_1)\mu},
\end{align}
which is just the symmetrization of $\tilde{T}^{\mu\nu\mu_1 \nu_1}$ with respect to $\mu\nu$, i.e.,
$T^{\mu\nu\mu_1 \nu_1} = \tilde{T}^{(\mu\nu)\mu_1 \nu_1}$.
Hence, $T^{\mu\nu\mu_1 \nu_1} (\ms^n,\ms^m)$ satisfies the following identity,
\begin{align}
T^{\mu\nu\mu_1 \nu_1} (\ms^n,\ms^m)=T^{\mu\nu\mu_1 \nu_1} (\ms^m,\ms^n).\label{gy11}
\end{align}
Furthermore, we should note the relation $\tilde{T}^{\mu\nu\mu_1 \nu_1} (\ms^m,\ms^m)=T^{\mu\nu\mu_1 \nu_1} (\ms^{m},\ms^m)$.

By using the identities (\ref{gy9}) and (\ref{gy11}), the dependence of Eqs.(\ref{gy5}), (\ref{gy6}), (\ref{gy7}) and (\ref{gy8}) on $\partial_{\mu}\partial_{\nu}g_{\alpha\beta}$ can be identified as follows,
\begin{align}
T_{(4m)} \sim& - \left\{ 
\sum_{k=0}^{2m-1} T^{\mu\nu\mu_1 \nu_1} (\ms^{k},\ms^{4m-k})
+ \frac12 T^{\mu\nu\mu_1 \nu_1} (\ms^{2m},\ms^{2m}) \right\} \partial_\mu \partial_\nu g_{\mu_1 \nu_1} ,\label{lll1}\\
T_{(4m+1)} 
\sim& -\sum_{k=0}^{2m} T^{\mu\nu\mu_1\nu_1} (\ms^k, \ms^{4m-k+1}) \partial_\mu \partial_\nu g_{\mu_1 \nu_1}, \label{lll2} \\
T_{(4m+2)} 
\sim&- \left\{ \sum_{k=0}^{2m}T^{\mu\nu\mu_1 \nu_1} (\ms^k, \ms^{4m-k+2})
+ \frac12 T^{\mu\nu\mu_1 \nu_1}(\ms^{2m+1},\ms^{2m+1}) \right\}\partial_\mu \partial_\nu g_{\mu_1 \nu_1}, \label{lll3} \\
T_{(4m+3)}\sim& -\sum_{k=0}^{2m+1}T^{\mu\nu\mu_1 \nu_1} (\ms^k,\ms^{4m-k+3})\partial_\mu \partial_\nu g_{\mu_1 \nu_1}. \label{lll4}
\end{align}
We should note that all the above terms (\ref{lll1}), (\ref{lll2}), (\ref{lll3}) and (\ref{lll4}) can be summarized as follows,
\begin{align}
T_{(K)}\sim-\frac12\sum_{k=0}^{K} T^{\mu\nu\mu_1 \nu_1} (\ms^k, \ms^{K-k})\partial_\mu \partial_\nu g_{\mu_1 \nu_1}, \label{iii2}
\end{align}
for $K=0,1,2,3\cdots$.
By substituting the above relation (\ref{iii2}) into (\ref{gy10}), 
we obtain,
\begin{align}
\nabla_\mu \left(\ms^{-1\nu}_{~~~~~\rho} \nabla_\mu Y^{\mu \rho}_{(n)} (\ms) \right)
=-\frac14 \sum_{K=0}^{n-1} (-1)^K
 e_{(n-K-1)} (\ms)\sum_{k=0}^{K} T^{\mu\nu\mu_1 \nu_1} (\ms^k, \ms^{K-k})\partial_\mu \partial_\nu g_{\mu_1 \nu_1}  + (\text{terms without }\partial_\mu \partial_\nu g_{\alpha \beta} ).
\end{align}
Finally, using the expansion formula (\ref{ssss1}) in Appendix \ref{ap1},
\begin{align}
Y_{(n-1)}^{\mu\nu(\mu_1\nu_1)}(\ms)
= \frac12 \sum_{K=0}^{n-1} (-1)^K e_{(n-K-1)} (\ms)\sum_{k=0}^{K} T^{\mu\nu\mu_1 \nu_1} (\ms^k, \ms^{K-k}),
\end{align}
we obtain the identity,
\begin{align}
\nabla_\nu \left(\ms^{-1\nu}_{~~~~~\rho} \nabla_\mu Y^{\mu \rho}_{(n)} (\ms) \right)
=-\frac12Y_{(n-1)}^{\mu\nu\mu_1\nu_1} (\ms) \partial_\mu \partial_\nu g_{\mu_1 \nu_1}  + (\text{terms without }\partial_\mu \partial_\nu g_{\alpha \beta} ).
\end{align}

\section{SUMMARY}
\label{summary}
We have shown the existence of an additional constraint of the dRGT model in the Lagrangian formulation and the metric formulation.
We found an identity (\ref{ly8}) which plays a crucial role for the existence of the additional constraint, and we have proved the identity for any dRGT potential terms.
This identity realizes the covariant expression of the additional constraint (\ref{covariant}) in a certain parameter region.

There is a possibility that the analysis in this paper may be extended to the Bimetric gravity \cite{Hassan3}, which is obtained by regarding $f_{\mu\nu}$ as a dynamical field and adding the Einstein-Hilbert term $M_{f}^{D-2} \int d^Dx \sqrt{-f}R(f)$ to the dRGT action (\ref{int2}). 
In \cite{bernard2}, the Bimetric gravity linearized around a general background solution has been considered.
They have deriven the linear combination of the linearized EoM which does not contain any second order time derivative terms of $\delta g_{0\mu}$ and $\delta f_{0\mu}$ (Here, we denote the perturbation of the metrics $g_{\mu\nu}, \ f_{\mu\nu}$ as $\delta g_{\mu\nu}, \ \delta f_{\mu\nu}$, respectively.).
This linear combination in the Bimetric gravity corresponds to the identity (\ref{ly8}).
Therefore, by extending the analysis in this paper to the Bimetric gravity, we may develop the tractable method for deriving the constraints in the Bimetric gravity.
%It may realize to express the constraints of the Bimetric gravity in the metric and the Lagrangian formulation to consider the nonlinear representation of the linear combination of the linearized EoM in \cite{bernard2}. 

\if0
In this paper, we considered only the constraint corresponding to $\partial^\mu h_{\mu\nu}=0, \ h=0$ in (\ref{inin1}). However, strictly speaking, there ought to be one more constraint from the point of view of the Lagrangian analysis which is just the Lagrangian version of the Hamilton analysis (see \cite{Buchbinder1,new curved,BTM} for the case of the FP model).
The additional constraint in the case of the FP model is obtained by taking the time derivative of $h=0$, i.e.
\begin{align}
\partial_0 h=0. \label{sum1}
\end{align}
This condition can be regarded as a constraint because it does not depend on second order time derivatives.
From the analysis in this paper, it is not clear that the existence of the full-nonlinear completion corresponding to (\ref{sum1}).
In order to prove this statement, we have to rewrite the constraint $\psi$ given in Eq.(\ref{sum2}) so that the dependence on $\partial_0 g_{\mu\nu}$ can be read off.
It may lead more simple formation of the constraint $\psi=0$ in Eq.(\ref{sum2}).
This is the future work.
\fi
\appendix

\section{Properties of $g^{\mu_1 \nu_1 \cdots \mu_n \nu_n}$}
\label{ap1}
\renewcommand{\theequation}{A.\arabic{equation}}
\setcounter{equation}{0}
In this appendix, we summarize the properties of the higher rank tensor $g^{\mu_1 \nu_1 \mu_2 \nu_2 \cdots \mu_n \nu_n}$.
\\
{\bf Definition:} Let us define the higher rank tensor $g^{\mu_1 \nu_1 \mu_2 \nu_2 \cdots \mu_n \nu_n}$ as follows,
\begin{align}
g^{\mu_1 \nu_1 \mu_2 \nu_2 \cdots \mu_n \nu_n} &\equiv 
n! \delta^{\nu_1}_{[\rho_1} \delta^{\nu_2}_{\rho_2} \cdots \delta^{ \nu_n }_{\rho_n]}
g^{\mu_1 \rho_1} g^{\mu_2 \rho_2} \cdots g^{\mu_n \rho_n}
\notag \\
 &=\frac{-1}{(D-n)!} E^{\mu_1 \mu_2 \cdots \mu_n \sigma_{n+1} \cdots \sigma_D } E^{\nu_1 \nu_2 \cdots \nu_n}_{~~~~~~~~~~\sigma_{n+1} \cdots \sigma_D }.  \label{Ap1}
\end{align}
Here, $E^{\mu_1 \nu_1 \cdots \mu_n \nu_n}$ is defined by using the Levi-Civita anti-symmetric tensor density $\epsilon^{\mu_1 \nu_1 \cdots \mu_n \nu_n}$ as follows,
\begin{align}
&E^{\mu_1 \mu_2 \cdots \mu_D} \equiv \frac{1}{\sqrt{-g}} \epsilon^{\mu_1 \mu_2 \cdots \mu_D} \notag \\
&\epsilon^{\mu_1 \mu_2 \cdots \mu_D}=
\begin{cases}
+1\text{~~~} (\mu_1 \mu_2 \cdots \mu_D) \text{ is an even permutation of } (0123\cdots ) \\
 -1\text{~~~} (\mu_1 \mu_2 \cdots \mu_D) \text{ is an odd permutation of } (0123\cdots ) \\
0 \; \mbox{~~~~~Otherwise}  
\end{cases}
\end{align}
\\
{\bf Example:} A few examples are given by,
\begin{align}
g^{\mu_1 \nu_1 \mu_2 \nu_2} \equiv& g^{\mu_1 \nu_1}g^{ \mu_2 \nu_2} - g^{\mu_1 \nu_2}g^{\mu_2 \nu_1}, \notag \\
g^{\mu_1 \nu_1 \mu_2 \nu_2 \mu_3 \nu_3} \equiv& 
g^{\mu_1 \nu_1} g^{\mu_2 \nu_2} g^{\mu_3 \nu_3}
+ g^{\mu_1 \nu_2} g^{\mu_2 \nu_3} g^{\mu_3 \nu_1}
+g^{\mu_1 \nu_3} g^{\mu_2 \nu_1} g^{\mu_3 \nu_2}\notag \\
&-g^{\mu_1 \nu_2} g^{\mu_2 \nu_1} g^{\mu_3 \nu_3}
-g^{\mu_1 \nu_1} g^{\mu_2 \nu_3} g^{\mu_3 \nu_2}
-g^{\mu_1 \nu_3} g^{\mu_2 \nu_2} g^{\mu_3 \nu_1}. \label{aaa1}
\end{align}
{\bf Symmetries:} The symmetries of $g^{\mu_1 \nu_1\mu_2 \nu_2 \cdots\mu_n \nu_n}$ can be summarized as follows, 
\begin{align}
&\mu_i \longleftrightarrow \mu_j \text{ :Anti-symmetric}, \notag \\
&\nu_i \longleftrightarrow \nu_j \text{ :Anti-symmetric}, \notag \\
&(\mu_i , \nu_i ) \longleftrightarrow (\mu_j , \nu_j ) \text{ :Symmetric}, \notag \\
&\{ \mu_i \} \longleftrightarrow \{ \nu_i \} \text{ :Symmetric}. \label{Appp1}
\end{align}
\\
{\bf Contraction:}
By using the expression of the second line of Eq.(\ref{Ap1}),
The contraction of $g^{\mu_1 \nu_1 \cdots \mu_n \nu_n}$ with respect to superscripts $\mu_n,\nu_n$ is proportional to
$2(n-1)$ th rank tensor $g^{\mu_1 \nu_1 \mu_{n-1} \nu_{n-1}}$,
\begin{align}
\label{Ap2}
{g^{\mu_1 \nu_1 \cdots \mu_{n-1} \nu_{n-1} \mu_n }}_{\mu_n}= (D-n+1) g^{\mu_1 \nu_1 \cdots \mu_{n-1} \nu_{n-1} } .
\end{align}
\\
{\bf Expansion:}
From the definition (\ref{Ap1}), 
a $2n$th-rank tensor can be expanded in products of a $2m$th-rank tensor and a $2(n-m)$th-rank tensor,
\begin{align}
g^{\mu_1 \nu_1 \cdots \mu_n \nu_n} &= \delta^{\nu_1 ~\nu_2 \cdots \nu_n}_{~\lambda _1~ \lambda_2 \cdots \lambda_n} 
g^{\mu_1 \lambda_1} \cdots g^{\mu_n \lambda_n} \notag \\
&= \delta^{\nu_1 ~\nu_2 \cdots \nu_n}_{~\lambda _1~ \lambda_2 \cdots \lambda_n} \frac{1}{m! (n-m)!}
 g^{\mu_1 \lambda_1 \cdots \mu_m \lambda_m} g^{\mu_{m+1} \lambda_{m+1} \cdots \mu_n \lambda_n }.    \label{Ap3}
\end{align}
A few examples are given by,
\begin{align}
g^{\mu_1 \nu_1 \mu_2 \nu_2 \mu_3 \nu_3} &= g^{\mu_1 \nu_1 } g^{\mu_2 \nu_2 \mu_3 \nu_3} + g^{\mu_1 \nu_2 } g^{\mu_2 \nu_3 \mu_3 \nu_1} 
+ g^{\mu_1 \nu_3} g^{\mu_2 \nu_1 \mu_3 \nu_2} , \notag \\
g^{\mu_1 \nu_1 \mu_2 \nu_2 \mu_3 \nu_3 \mu_4 \nu_4} &= g^{\mu_1 \nu_1 } g^{\mu_2 \nu_2 \mu_3 \nu_3 \mu_4 \nu_4} -
 g^{\mu_1 \nu_2 } g^{\mu_2 \nu_1 \mu_3 \nu_3 \mu_4 \nu_4} 
- g^{\mu_1 \nu_3} g^{\mu_2 \nu_2 \mu_3 \nu_1 \mu_4 \nu_4} - g^{\mu_1 \nu_4} g^{\mu_2 \nu_2 \mu_3 \nu_3 \mu_4 \nu_1}.
\end{align}
\\
{\bf 1+(D-1) decomposition:}
 Following formula is also useful,
\begin{align}
g^{00\mu_1 \nu_1 \mu_2 \nu_2 \cdots \mu_n \nu_n }
= g^{00} \theta^{\mu_1 \nu_1 \mu_2 \nu_2 \cdots \mu_n \nu_n}.\label{klkl1}
\end{align}
Here, by using $\theta^{\mu\nu}$ which is the projection operator living in $D-1$ space orthogonal to time-like direction $g^0_\mu$, $\theta^{\mu_1 \nu_1 \mu_2 \nu_2 \cdots \mu_n \nu_n}$ is defined as follows,
\begin{align}
&\theta^{\mu_1 \nu_1 \mu_2 \nu_2 \cdots \mu_n \nu_n} \equiv n! \delta^{\nu_1}_{[\rho_1} \delta^{\nu_2}_{\rho_2} \cdots \delta^{\nu_n}_{\rho_n]} \theta^{\mu_1 \rho_1}
\theta^{\mu_2 \rho_2} \cdots \theta^{\mu_n \rho_n}, \notag \\
&\theta^{\mu\nu} \equiv g^{\mu\nu} - \frac{g^{0\mu} g^{0\nu}}{g^{00}}.
\end{align}
\\
{\bf Expansion of $Y^{\mu\nu}_{(n)}$:}
Let us define the matrices $Y_{(n)}^{\mu\nu}(\ms)$ as follows,
\begin{align}
Y^{\mu\nu}_{(n)}(\mathcal{S}) \equiv \frac{1}{n!} g^{\mu\nu\mu_1 \nu_1 \mu_2 \nu_2 \cdots \mu_n \nu_n} \mathcal{S}_{\mu_1 \nu_1}
\mathcal{S}_{\mu_2 \nu_2} \cdots \mathcal{S}_{\mu_n \nu_n}.
\end{align}
Although the tensor $\ms_{\mu\nu}$ denotes the square root matrix given in (\ref{int2}), 
the following properties are valid for any symmetric matrices $\ms_{\mu\nu}=\ms_{(\mu\nu)}$.
$Y^{\mu\nu}_{(n)}(\ms)$ can be expanded in powers of $\left[ \ms^n\right]^{\mu\nu}$ as follows,
\begin{align}
Y^{\mu\nu}_{(n)} (\ms)= \sum_{k=0}^{n} (-1)^k e_{(n-k)} (\ms ) \left[\ms^k \right]^{\mu\nu},  
\label{ky1}
\end{align}
This relation can be obtained by solving the recursion relation,
\begin{align}
Y^{\mu\nu}_{(n)}(\ms) = g^{\mu\nu}e_{(n)}(\ms) - \ms^\mu_{~\rho} Y^{\rho \nu}_{(n-1)} (\ms),
\label{ly3}
\end{align}
which can easily be shown by using the following expansion relation,
\begin{align}
\delta^{\mu_1~~\mu_2~~\mu_3\cdots \mu_n}_{~~~\nu_1~~\nu_2 ~~\nu_3\cdots \nu_n}
=n\delta^{\mu_1}_{~~~[\nu_1}\delta^{\mu_2~~\mu_3\cdots \mu_n}_{~~~\nu_2~~\nu_3\cdots \nu_n]}.
\end{align}
\\
{\bf Expansion of $Y^{\mu\nu (\mu_1 \nu_1)}_{(n)}$:}
Furthermore, we derive a expansion formula for the quantity $Y^{\mu\nu \mu_1 \nu_1}_{(n-1)}(\ms)$ defined as follows,
\begin{align}
Y^{\mu\nu \mu_1 \nu_1}_{(n-1)} (\ms)\equiv  \frac{1}{(n-1)!} g^{\mu\nu\mu_1 \nu_1 \mu_2 \nu_2 \cdots \mu_n \nu_n} 
\mathcal{S}_{\mu_2 \nu_2} \cdots \mathcal{S}_{\mu_n \nu_n}.
\end{align}
The symmetrized quantity $Y^{\mu\nu(\mu_1 \nu_1)}_{(n)}(\ms)$ can be expanded as follows,
\begin{align}
 Y_{(n-1)}^{\mu\nu(\mu_1\nu_1)}(\ms)
&=\sum_{K=0}^{n-1} (-1)^K e_{(n-K-1)} (\ms)\sum_{k=0}^{K}
\left[ \left[ \ms^{K-k}  \right]^{\mu_1 \nu_1} \left[\ms^k \right]^{\mu\nu}
-\left[ \ms^k \right]^{\mu(\mu_1} \left[\ms^{K-k} \right]^{\nu_1)\nu}  \right]\notag \\
&= \frac12 \sum_{K=0}^{n-1} (-1)^K e_{(n-K-1)} (\ms)\sum_{k=0}^{K} T^{\mu\nu\mu_1 \nu_1} (\ms^k, \ms^{K-k})
.\label{ssss1}
\end{align}
The proof is given by,
\begin{align}
Y_{(n-1)}^{\mu\nu(\mu_1\nu_1)}(\ms)
=& \frac{\partial Y^{\mu\nu}_{(n)}(\ms)}{\partial \ms_{\mu_1 \nu_1}} \notag \\
=&\sum_{k=0}^{n-1} (-1)^k  \sum_{l=0}^{n-k-1} (-1)^l e_{(n-k-1-l)} (\ms)
\left[ \ms^l \right]^{\mu_1 \nu_1} \left[\ms^k\right]^{\mu\nu} \notag \\
&+\sum_{k=1}^{n}(-1)^k e_{(n-k)}(\ms) \sum_{l=0}^{k-1} \left[ \ms^l \right]^{\mu(\mu_1}
\left[ \ms^{k-l-1}\right]^{\nu_1)\nu}. \label{ssss2}
\end{align}
Here, we use the expansion formula (\ref{ky1}).
By define a new index as $K\equiv k+l$, we can show,
\begin{align}
\sum_{k=0}^{n-1} (-1)^k  \sum_{l=0}^{n-k-1} (-1)^l e_{(n-k-1-l)} (\ms)
\left[ \ms^l \right]^{\mu_1 \nu_1} \left[\ms^k\right]^{\mu\nu}
=\sum_{K=0}^{n-1} \sum_{k=0}^{K} (-1)^K e_{(n-K-1)}(\ms)
 \left[ \ms^{K-k}\right]^{\mu_1 \nu_1} \left[ \ms^k \right]^{\mu\nu}. \label{ssss3}
\end{align}
By substituting (\ref{ssss3}) into (\ref{ssss2}), we obtain (\ref{ssss1}).

\section{Detailed calculations}
\label{ap2}
\renewcommand{\theequation}{B.\arabic{equation}}
\setcounter{equation}{0}
In this appendix, we summarize the detailed calculation of Sec.\ref{pro}.
\\{\bf Derivation of (\ref{gy4}) {and} (\ref{hy6}):} {The derivations of Eqs.(\ref{gy4}) {and} (\ref{hy6}) are given by,}
\begin{align}
\ms^{-1\nu}_{~~~~~\rho} \nabla_{\mu} \left[\ms^{4m+1} \right]^{\mu\rho}
&=\ms^{-1\nu}_{~~~~~\rho}  \left[\nabla_{\mu}\ms^{2m}\ms^{2m+1}+
\ms^{2m+1}\nabla_{\mu}\ms^{2m} + \ms^{2m} \nabla_\mu \ms \ms^{2m} \right]^{\mu\rho} 
\notag \\
&= \left[ \left( \nabla_{\mu}\ms^{2m}\ms^{2m+1}+
\ms^{2m+1}\nabla_{\mu}\ms^{2m} \right)\ms^{-1}
+ \ms^{2m} \nabla_\mu \ms \ms^{2m-1} \right]^{\mu\nu} 
\notag \\
&= \left[ \left( \nabla_{\mu}f^{m}\ms^{2m+1}+
\ms^{2m+1}\nabla_{\mu}f^{m} \right)\ms^{-1}
+ \frac12 \ms^{2m-1} \nabla_\mu f \ms^{2m-1} \right]^{\mu\nu}
+ \left[ \ms^{2m} \nabla_\mu \ms \ms^{2m-1} \right]^{[\mu\nu]}, \label{qqq1} \\
\ms^{-1\nu}_{~~~~~\rho} \nabla_{\mu} \left[\ms^{4m+3} \right]^{\mu\rho}
&=\ms^{-1\nu}_{~~~~~\rho}  \left[\nabla_{\mu}\ms^{2m+2}\ms^{2m+1}+
\ms^{2m+1}\nabla_{\mu}\ms^{2m+2} - \ms^{2m+1} \nabla_\mu \ms \ms^{2m+1} \right]^{\mu\rho} 
\notag \\
&= \left[ \left( \nabla_{\mu}\ms^{2m+2}\ms^{2m+1}+
\ms^{2m+1}\nabla_{\mu}\ms^{2m+2} \right)\ms^{-1}
- \ms^{2m+1} \nabla_\mu \ms \ms^{2m} \right]^{\mu\nu} 
\notag \\
&= \left[ \left( \nabla_{\mu}f^{m+1}\ms^{2m+1}+
\ms^{2m+1}\nabla_{\mu}f^{m+1} \right)\ms^{-1}
- \frac12 \ms^{2m} \nabla_\mu f \ms^{2m} \right]^{\mu\nu}
 -\left[ \ms^{2m+1} \nabla_\mu \ms \ms^{2m} \right]^{[\mu\nu]}.
\label{qqq2}
\end{align}
{Here, we use the Eq.(\ref{hy5}) in the last lines of each Eqs.(\ref{qqq1}) {and} (\ref{qqq2}).} 
\\
{\bf Derivation of (\ref{gy5}) {and} (\ref{gy7}):}
In the case of $K=4m$, (\ref{gy1}) becomes,
\begin{align}
T_{(4m)} \equiv \sum_{k=0}^{4m} \left[ \ms^{k-1} \right]^{\mu\nu} \left[ \ms^{4m-k-1} \right]^{\mu_1 \nu_1} \nabla_\nu \nabla_\mu f_{\mu_1\nu_1} 
-2 \nabla_\nu \left( \ms^{-1\nu}_{~~~~~\rho} \nabla_\mu \left[ \ms^{4m+1} \right]^{\mu \rho}\right). \label{fy1}
\end{align}
The coefficient operator of the first term can be decomposed as follows,
\begin{align}
\sum_{k=0}^{4m} \left[ \ms^{k-1} \right]^{\mu\nu} \left[ \ms^{4m-k-1} \right]^{\mu_1 \nu_1} 
=& \sum_{k'=0}^{2m} \left[ \ms^{2k'-1} \right]^{\mu\nu} \left[ \ms^{4m-2k'-1} \right]^{\mu_1 \nu_1} 
+ \sum_{k'=0}^{2m-1}   \left[ \ms^{2k'} \right]^{\mu\nu} \left[ \ms^{4m-2k'-2} \right]^{\mu_1 \nu_1} \notag \\
=&\sum_{k'=0}^{m-1} \left[  \left[ \ms^{2k'-1} \right]^{\mu\nu} \left[ \ms^{4m-2k'-1} \right]^{\mu_1 \nu_1}+\left[ \ms^{4m-2k'-1} \right]^{\mu\nu} \left[ \ms^{2k'-1} \right]^{\mu_1 \nu_1}   \right] \notag \\
&+ \sum_{k'=0}^{m-1}  \left[ \left[ \ms^{2k'} \right]^{\mu\nu} \left[ \ms^{4m-2k'-2} \right]^{\mu_1 \nu_1} +  \left[ \ms^{4m-2k'-2} \right]^{\mu\nu} \left[ \ms^{2k'} \right]^{\mu_1 \nu_1}  \right] \notag \\
&+ \left[ \ms^{2m-1} \right]^{\mu\nu} \left[ \ms^{2m-1} \right]^{\mu_1 \nu_1}.\label{fy2}
\end{align}
In the first line, we separate the set of $k$ into even numbers ${k=}2k'$ and odd numbers ${k=}2k'+1$.
In the second line, we separate the region $0\le k' \le 2m-1$ into $0\le k' \le m-1$ and $m\le k'\le 2m-1$, and redefine the indexes.
On the other hand, by using the formula (\ref{gy4}), the second term of (\ref{fy1}) can be expressed as follows,
\begin{align}
\nabla_\nu \left( \ms^{-1\nu}_{~~~~~\rho} \nabla_\mu \left[ \ms^{4m+1} \right]^{\mu \rho} \right)
\sim &\left\{ \sum_{k=0}^{m-1} \left[ \left[ \ms^{4m-2k-1} \right]^{\mu (\mu_1}\left[ \ms^{2k-1} \right]^{\nu_1)\nu} 
+\left[ \ms^{2k} \right]^{\mu(\mu_1} \left[ \ms^{4m-2k-2} \right]^{\nu_1) \nu} 
\right] \right. \notag \\
&\left. +\frac12 \left[ \ms^{2m-1} \right]^{\mu(\mu_1} \left[ \ms^{2m-1} \right]^{\nu_1) \nu}
\right\} \nabla_\nu \nabla_\mu f_{\mu_1 \nu_1}
+\left[ \ms^{2m}\nabla_\nu \nabla_\mu \ms \ms^{2m-1}\right]^{[\mu\nu]}. \label{fy3}
\end{align}
Here, we use the {relation} ``$\sim$" which means the equivalence up to terms without $\partial_\mu \partial_\nu g_{\alpha \beta}$.
By substituting (\ref{fy2}) and (\ref{fy3}) into (\ref{fy1}), we obtain (\ref{gy5}).
As a similar way, we can derive (\ref{gy7}) by using (\ref{hy6}).
\\
{\bf Derivation of (\ref{gy6}) {and} (\ref{gy8}):}
Let us derive Eq.(\ref{gy6}).
In the case of $K=4m+1$,  (\ref{gy1}) becomes,
\begin{align}
T_{(4m+1)} \equiv \sum_{k=0}^{4m+1} \left[ \ms^{k-1} \right]^{\mu\nu} \left[ \ms^{4m-k} \right]^{\mu_1 \nu_1} \nabla_\nu \nabla_\mu f_{\mu_1\nu_1} 
-2 \nabla_\nu \left( \ms^{-1\nu}_{~~~~~\rho} \nabla_\mu \left[ \ms^{4m+2} \right]^{\mu \rho}\right).
\label{uy1}
\end{align}
As we have derived Eq.(\ref{fy2}), the first term of the right hand side of the above equation (\ref{uy1}) can be deformed as,
\begin{align}
\sum_{k=0}^{4m+1} \left[ \ms^{k-1} \right]^{\mu\nu} \left[ \ms^{4m-k} \right]^{\mu_1 \nu_1}
=&\sum_{k'=0}^{m-1} \left[  \left[ \ms^{2k'-1} \right]^{\mu\nu} \left[ \ms^{4m-2k'} \right]^{\mu_1 \nu_1}+\left[ \ms^{4m-2k'} \right]^{\mu\nu} \left[ \ms^{2k'-1} \right]^{\mu_1 \nu_1}   \right] \notag \\
&+ \sum_{k'=0}^{m}  \left[ \left[ \ms^{2k'} \right]^{\mu\nu} \left[ \ms^{4m-2k'-1} \right]^{\mu_1 \nu_1} +  \left[ \ms^{4m-2k'-1} \right]^{\mu\nu} \left[ \ms^{2k'} \right]^{\mu_1 \nu_1}  \right].\label{uy2}
\end{align}
On the other hand, the second term of the right hand side of Eq.(\ref{uy1}) can be expressed as,
\begin{align}
\nabla_\nu \left( \ms^{-1\nu}_{~~~~~\rho} \nabla_\mu \left[ \ms^{4m+2} \right]^{\mu \rho}\right)
\sim&
\sum_{k=0}^{2m} \left[ \ms^{2k-1} \nabla_\nu \nabla_\mu f  \ms^{4m-2k}  \right]^{\nu\mu} \notag \\
=& \sum_{k=0}^{m-1} \left[\ms^{2k-1} \nabla_\nu \nabla_\mu f \ms^{4m-2k} \right]^{\nu\mu}
+\sum_{k=0}^{m} \left[ \ms^{4m-2k-1} \nabla_\nu \nabla_\mu  f \ms^{2k} \right]^{\nu\mu}.\label{uy3}
\end{align}
In the second line, we separated the region $0\le k \le 2m$ into $0\le k \le m-1$ and $m\le k \le 2m$, and redefined the indexes.
By substituting (\ref{uy2}) and (\ref{uy3}) into (\ref{uy1}), we obtain Eq.(\ref{gy6}).
As a similar way, we obtain Eq.(\ref{gy8}).

\end{document}